\documentclass[12pt]{iopart}

\usepackage[english]{babel}
\usepackage{multirow}
\usepackage{pifont}                           
\usepackage{amssymb}                          
\usepackage{graphicx}    
\usepackage{subfig}                      
\usepackage{bm}                                
\usepackage{color}
\usepackage{psfrag}
\usepackage{fix-cm}
\usepackage{cite}

\begin{document}

\title[]{Engineering Dark Solitary waves in Ring-Trap Bose-Einstein condensates}

\author{D.\ Gallucci and N.\ P.\ Proukakis}

\address{Joint Quantum Centre (JQC) Durham-Newcastle, School of Mathematics 
and Statistics, Newcastle University, Newcastle upon Tyne, NE1 7RU, 
United Kingdom
}



\begin{abstract}

We demonstrate the feasibility of generation of quasi-stable counter-propagating solitonic structures in an atomic Bose-Einstein condensate confined in a realistic toroidal geometry, and identify optimal parameter regimes for their experimental observation.
Using density engineering we numerically identify distinct regimes of motion of the emerging macroscopic excitations, including both solitonic motion along the azimuthal ring direction, such that structures remain visible after multiple collisions even in the presence of thermal fluctuations, and snaking instabilities leading to the decay of the excitations into vortical structures.
Our analysis, which considers both mean field effects and fluctuations, is based on the ring trap geometry of Murray {\it et al.} 2013
\textit{Phys. Rev. A} {\bf 88} 053615.

\end{abstract}

\maketitle

\section{Introduction \label{Introduction}}

The emerging field of atomtronics \cite{seaman2007,pepino2009} is associated with the creation of atomic circuit architectures based on ultracold atoms. A promising candidate for a closed prototype atomtronic circuit is based on laser-beam manipulation of ultracold atoms confined in toroidal geometries \cite{Eckel2014.nature}, a situation readily available in numerous laboratories 
\cite{Arnold2006,Ryu2007,Ramanathan2011,Wright2013.prl,Wright2013.pra,Eckel2014.prx,Jendrzejewski2014,Eckel2014.nature,
Henderson2009,Moulder2012,Corman2014,Heathcote2008,Sherlock2011,Gupta2005,Sauer2001,Lesanovsky2007,Mason2009,halkyard2010,Aftalion2010}.
Harnessing such circuits for technological applications (e.g. rotation sensors) requires a detailed understanding of the dynamics induced in such geometries through controlled perturbations, which has recently become very timely.
Parallel to this, nonlinear excitations in the form of solitons could be useful for potential applications, e.g. due to their repetitive motion in a closed circuit and robustness against collisions.

The aim of this work is to demonstrate that although there are no known stable azimuthal solitonic solutions in toroidal geometries (somewhat related radial excitations in the form of `ring dark solitons' have been discussed in \cite{kevrekidis_frantzeskakis_book_08,Theocharis2003,Toikka2013,Toikka2014,Toikka2014b,wang2015}), soliton-like structures propagating at a fraction of the speed of sound and largely maintaining their shape after numerous collisions can nonetheless be generated through density engineering, even in the presence of thermal fluctuations.

In the context of ultracold atoms, solitonic nonlinear excitations arise spontaneously at the phase 
transition, as a consequence of quenching the system from the thermal to the 
condensed regimes through the Kibble-Zurek 
mechanism \cite{kibble_76,zurek_85,Lamporesi2013,Zurek2009,Weiler2008}, 
or can be engineered by means of 
well known techniques such as phase imprinting 
\cite{Burger2002,Dobrek1999,Denschlag2000}, density engineering  
\cite{Gredeskul1990,Dutton2001,Burger2002,Brazhnyi2003}, or a combination of 
both \cite{Carr2001,Burger2002}.
The creation of solitons by engineering the density of the gas is 
typically performed by using a blue-detuned laser beam focussed in a narrow region 
of the system, on the scale of the healing length. The density 
distribution of the gas adapts to the presence of this perturbation, and the atoms 
are repelled from the region where the laser field is applied. Imposing a sharp 
density feature in a Bose-Einstein condensate determines a localised dip in the distribution 
which should then lead to the generation of solitons upon removal of the laser field. 
Solitons are one-dimensional objects originating from a unique balance between the 
kinetic energy, associated with spatial variations of 
the order parameter, and the atom-atom interaction energy; these waves largely preserve  
their shape after colliding with each other (undergoing only a phase shift).
Although the initial engineered dark soliton experiments in harmonic traps led to both dynamical 
\cite{Carr2000,Feder2000,Anderson2001,Muryshev2002,brand2002,Proukakis2004.job,Parker2010,Shomroni2009} and thermal \cite{Burger1999,Muryshev1999,Muryshev2002,Jackson2007} instabilities, both long-lived \cite{becker_stellmer_08} and stable \cite{weller_ronzheimer_08,Nguyen2014} solitons can now be routinely engineered in the lab.

In this work we demonstrate that long-lived structures resembling one-dimensional dark solitons can also be engineered as counter-propagating pairs in ring-shaped traps within appropriate parameter regimes and excitation schemes, and thus study their stability, dynamics and interactions. More specifically, such structures are generated here numerically via the density engineering scheme, based on the (gradual)
addition and removal of a Gaussian perturbation on a trapped Bose-Einstein condensate.
To achieve optimal dynamical stability of such structures, one would need to restrict investigations to a very idealised regime of tight confinement in {\em both} transverse and radial directions, which effectively reduces the system to the one-dimensional (1D) regime. Given the current significant experimental challenges in reaching this idealised 1D ring-trap regime, a pertinent question relates to how far from this regime one can deviate before dynamical instabilities dominate, severely limiting or even prohibiting solitonic behaviour, with a related question arising on the destabilizing role of thermal fluctuations.
Here, we demonstrate the existence of a broad experimentally-relevant regime, where such engineered structures remain relatively robust both against dynamical and thermal instabilities, also surviving through multiple collisions.
Our analysis highlights both the role of geometry and temperature in the evolution of such emerging solitary waves. 

After discussing the system geometry and identifying relevant ``control parameters'' (Sec.\,\ref{parameters}), we focus on the question of optimisation of the generation of such solitary waves, by means of the Gross-Pitaevskii equation (Sec.\,\ref{zero_temperature}).
Having identified optimum generation schemes, we then investigate the extent to which such structures could be obtained under realistic experimental conditions, in the presence of thermal fluctuations included here through numerical simulations of the Stochastic Gross-Pitaevskii equation \cite{Stoof2001,Gardiner2003} (Sec.\,\ref{finite_temperature}).
The latter approach has already been demonstrated as an excellent model for
 \textit{ab initio} 
equilibrium predictions of six independent quasi-2D \cite{Cockburn2012} and quasi-1D 
\cite{Cockburn2011a,Gallucci2012} Bose gas experiments,
and has also been used to investigate condensate growth \cite{Stoof2001,Proukakis2003_lasphys} and dark soliton dynamics in quasi-1D 
geometries \cite{Cockburn2010,Cockburn2011}, with the closely related Stochastic Projected Gross-Pitaevskii Equation \cite{blakie2008,blakie2008b,rooney2012,Gardiner2003} used to study spontaneous defect formation following a quench \cite{Weiler2008,Gary2014,McDonald2015}, vortex dynamics \cite{Simula2006}, and decay of persistent currents \cite{Rooney2013}. Further details of dynamics following a non-optimal choice of density engineering parameters and a comparison to the idealised 1D regime are discussed in two Appendices.

\section{Physical Set-up and Parameter regime \label{parameters}}

\begin{figure}[t]
\centering
\includegraphics[scale=0.6]{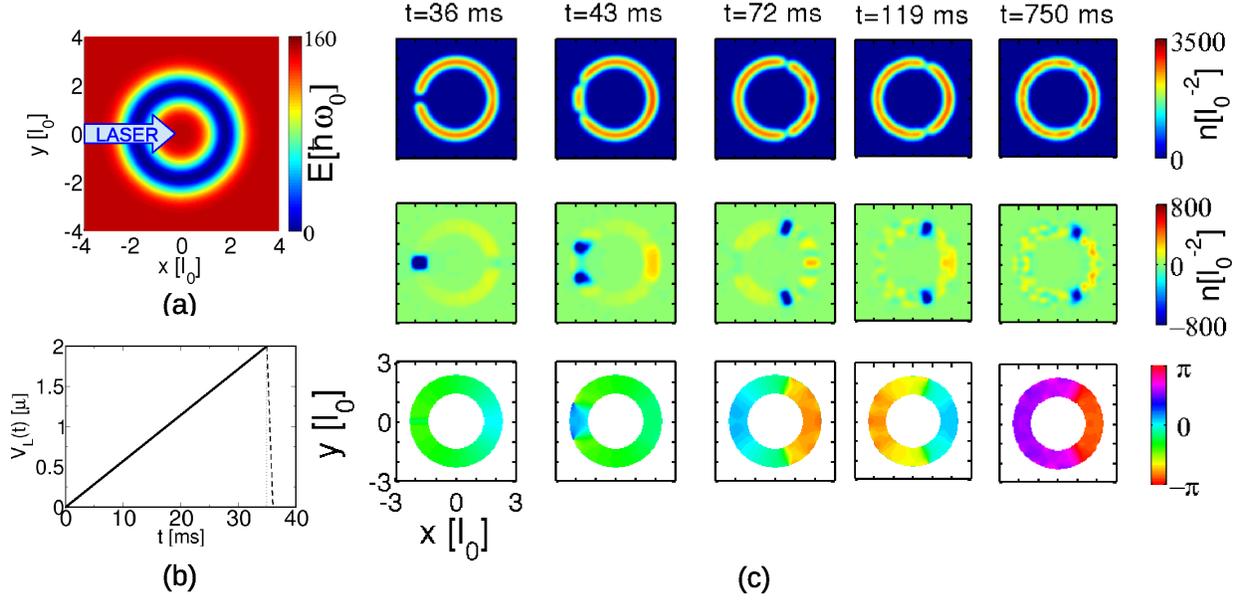}
  \caption{(a) Schematic of the 2D 
  ring-shaped confining potential (defined by Eq.\,(\ref{eq_ringtrap})), showing also where the blue-detuned laser is applied (see main text). 
  (b) Evolution of the amplitude of the imposed perturbing potential in units of the chemical potential $\mu$; the perturbing potential is  linearly ramped up to 
  $V_0=2\,\mu$ (solid line) and down to $V_0=0$ (dashed line) over a 
  time period of $35.5\,{\rm m}s$ and $0.5\,{\rm m}s$ respectively, as defined by Eq.~(3); a vertical 
  dotted line is shown here for reference, in order to distinguish our 
  procedure from a sudden turning off of the barrier.
 (c) Density (top), renormalized `carpet' (middle) and phase (bottom) 
  plots at times $t=36,43,72,119,750\,{\rm{ms}}$ (left to right) after switching 
  on the perturbation, with $t=36$ms corresponding to the time when the perturbation has just been turned off. The renormalized `carpet' plots are obtained in the usual way, by subtracting from the perturbed instantaneous density the
static density profile prior to the addition of the perturbation. The emergence of `solitonic' excitations is evident from the combined density and phase information with $t=119$ and $750$\,ms respectively corresponding to the cases after one and thirteen collisions, thus demonstrating that the generated `solitonic' structures remain largely unaffected by multiple collisions. To hide spurious features 
  in the phase plots, a mask has been used where the density is lower than $10\%$ 
  of the peak density at equilibrium.
[Parameters: $N=15625$, $\sigma/\xi\approx 0.7$, $l_{\rm r}/\xi=1.3$, $V_0/\mu=2$, with $\xi=1.5\,{\rm \mu m}$, such that we are probing the 2D solitonic regime $l_\perp<\xi<l_{\rm r}$; corresponding 2D peak density $\approx 25$ atoms per $\mu m^2$.]
  }
\label{initial_fig}.
\end{figure}

We consider a trapped ultracold atomic gas ($^{23}$Na atoms, scattering length $a_s=2.75 {\rm nm}$) 
confined in a ring-shaped trap of the form (see Fig.\,\ref{initial_fig}a for a 
visual representation):
\begin{equation}
\quad\quad\quad\quad\quad\quad
V({\bf r})=V_{G}\big(1-e^{-2({\rm r}-{\rm r_0})^2/w^2}\big)
\label{eq_ringtrap}
\end{equation}
where $V_{G}$, $r_0$ and $w$ are respectively the depth, radius and 
$1/e^2$ half-width of the ring-gaussian potential. The radius, $r_0$, is the distance 
from the center, where the potential reaches its minimum, whereas a length of $2w$ specifies the effective size of the ring channel where 
the gas is confined.

We assume tight transverse harmonic confinement  
with frequency $\omega_\perp$ in the direction perpendicular to the $(x,y)$ plane, such that, for sufficiently small but realistic atom numbers, the gas can be brought into the transverse ground 
state, thus reducing all system dynamics to effectively  two-dimensional.

We attempt to generate solitons via the density engineering 
technique, as this appears to be most relevant to recent experimental efforts \cite{priv_comm_phillips}. Specifically we envisage 
perturbing an initial equilibrium density for a (quasi-2D) ring filled with a BEC by gradually ramping 
on the intensity of a blue-detuned laser sheet focussed in a localised region of the gas, and subsequently removing the laser. 
This method is modelled by adding (to the ring-trap of Eq.\,(\ref{eq_ringtrap})) 
a narrow Gaussian potential of the form:
\begin{equation}
\quad\quad\quad\quad\quad\quad V_{\rm pert}({\bf r},t)=V_{\rm L}(t)\,e^{-y^2/2\sigma^2}\left( 1 - \Theta(x) \right) \;,
\label{eq_perturbation}
\end{equation}
i.e.\ applied in the left half-plane (region across $y=0$, for $x<0$, 
see Fig.\,\ref{initial_fig}a for reference), where $\Theta(x)$ is the step function,
$\sigma$ is the half-width 
of the Gaussian barrier and $V_{\rm L}(t)$ the time-dependent laser amplitude.
The ramp needs to be kept on for a time long enough (few ${\rm ms}$ -- tens of ${\rm ms}$) to create a sufficiently deep notch 
in the density distribution, which should then lead to the generation of a (single) 
pair of counter-propagating solitary waves.

In order to minimize the linear (sound wave) excitations emerging from sudden perturbations, we follow a scheme similar to that used in the experimental work of Ref.~\cite{Shomroni2009}, such that
the perturbing potential is linearly ramped up over a time $\tau_{\rm on}$ to its maximum value, $V_0$, 
and then ramped down to $0$ 
according to (perturbation on for times $0\le t \le \tau_{\rm pert}$):
\begin{eqnarray}
V_{\rm L}(t) &=& V_0 \left(\frac{t}{\tau_{\rm on}}\right)\left[\Theta(t)-\Theta(t-\tau_{\rm on})\right] \nonumber \\
&+& V_0 \left(\frac{1}{\tau_{\rm pert}-\tau_{\rm on}}\right) \left( \tau_{\rm pert}-t \right) \left[ \Theta(t - \tau_{\rm on}) - \Theta(t-\tau_{\rm pert})\right] \;,
\end{eqnarray}
where $\tau_{\rm on}$ the duration of the ramping-on sequence and $\left( \tau_{\rm pert}-\tau_{\rm on} \right)$ the ramping off timescale, which we have chosen to be relatively short. 

Consistent with Ref.~\cite{Shomroni2009} we find that adiabatically ramping up the perturbation limits the sound emitted, thus leading to `clean' profiles, without compromising the depth (or, equivalently, speed) of the emerging macroscopic excitations, which depends on the maximum value of $V_0(t)$. Although one could have used a slightly  smoother (e.g. parabolic) turning on/off ramp to give less perturbation, our linear scheme  seems to offer a sufficient reduction in background density noise facilitating our subsequent analysis. 
Thus, throughout this work, we show results for $\tau_{\rm on} = 35.5$ms and a ramping down time $\left( \tau_{\rm pert} - \tau_{\rm on} \right) = 0.5$ms.
The particular shape of the perturbation is shown for the chosen parameters in Fig.\,\ref{initial_fig}b. 

The underlying system geometry and induced perturbation lead to 
three physically-distinct sets of controllable parameters, respectively characterising:
(i) the unperturbed ring trap potential $V({\bf r})$ (depth $V_{G}$, location of minimum of confined potential $r_0$, width of ring $w$) with the transverse frequency $\omega_\perp$ entering our analysis implicitly, through its contribution to the effective two-dimensional interaction parameter, $g_{_{\rm 2D}} \propto \sqrt{\omega_\perp}$; (ii) the density-engineering perturbation $V_{\rm pert}$ 
(amplitude $V_0$, width $\sigma$, duration and slope of its application); and 
(iii) the details of the confined gas (atomic species, atom number $N$, characterised through the two-dimensional chemical potential $\mu$, $s$-wave scattering length $a_s$, and temperature $T$). 
Although this cumulatively leads to a very broad parameter diagram to be probed,
given a particular physical configuration and excitation scheme, the main physics is actually set only by a few parameters (or rather, their ratios).

To demonstrate this we choose to work with the particular experimental ring trap geometry of Ref.\,\cite{Murray2013}, namely: $\omega_\perp= 2 \pi \times 600\,\rm{Hz}$, $r_0=18.5\,\rm{\mu m}$, 
$w=9.45\,\rm{\mu m}$ and $V_G/k_{\rm B}=31.5\,\rm{nK}$ (Notice that in Sec.~\ref{finite_temperature} we work in the temperature range $1\,{\rm nK}<T<10\,{\rm nK}$, i.e. $0.04\,{\rm nK}<k_{\rm B}T/\hbar\omega_\perp<0.35\,{\rm nK}$). 
This, in turn, fixes the radial harmonic oscillator length, $l_{\rm r}=\sqrt{\hbar/m\omega_r}=2\,{\mu}m$
(where $\omega_r=\sqrt{4V_G/mw^2}$) and the transverse spatial extent, $l_{\perp}=\sqrt{\hbar/m\omega_\perp} = 0.86\,{\mu}m$.
Following Ref.\,\cite{Murray2013}, we introduce here a reference `length unit' 
$l_0=\sqrt{(\hbar/m\omega_0)}=10\,{\rm{\mu m}}$ (using $\omega_0/(2\pi)=4.4\,{\rm Hz}$), to which all our results are scaled.

To maintain a quasi-2D geometry, suppressing instabilities due to coupling to dynamics outside the $(x,y)$ plane, we work here with an atom number $N \approx 15000$\, (which leads to a corresponding peak density of $2500\,l_0^{-2}$ corresponding roughly to 25 atoms$/{\mu m}^2$). This choice ensures that the 2D condition  $\mu<\hbar \omega_\perp$ holds:
specifically, we choose a chemical potential 
$\mu \sim \hbar \omega_\perp / 3$.
For such an atom number we are typically in the 2D regime defined by $l_\perp < \xi < l_{\rm r}$, although a further dimensional reduction to a 1D regime ($l_\perp<l_{\rm r}<\xi$) is theoretically feasible through a slight reduction (by a factor of 2-3) of the scattering length by means of Feshbach resonances.

To ensure all atoms remain confined within the ring trap, we also restrict the system temperature, $T$, to values sufficiently below $V_{\rm G}/k_{\rm B}$.
In general, the density engineering method can lead to the generation of one (or more) solitons \cite{Burger2002}. To simplify the dynamics and avoid the generation of multiple pairs of counter-propagating structures, we thus choose the width 
of the perturbation $\sigma \approx \xi$ where $\xi$ is the minimum value of the healing length, as calculated at the peak density
\cite{Burger2002}.

Increasing the width of the perturbation $\sigma$ to values $\sigma/\xi>1$ (or significantly increasing $V_0$ for a fixed value of $\sigma$) leads to the generation of more than one pair of counter-propagating solitary waves, as discussed in  Appendix A.

Based on the choices described above, this effectively leaves us with the more manageable task of only 3 control parameters affecting the generation and subsequent propagation of the nonlinear excitations:

\begin{itemize}

\item The healing length of the system, $\xi$, broadly parametrizing the spatial extent of the emerging macroscopic excitation (e.g.\ 
dark soliton or vortex): this is defined here as $\xi=\hbar/\sqrt{(mg_{_{\rm 2D}}n_0)}$,
where $g_{_{\rm 2D}}=\sqrt{8\pi}(a_{\rm s}/l_\perp)(\hbar^2/m)$ is the 2D 
interaction strength, $a_{\rm s}$ is the scattering length and $n_0$ refers to the maximum density.
For a fixed transverse confinement $\omega_{\perp}$ investigated here, the value of $\xi$ can be controlled either by changing the number of atoms 
in a given geometry, which affects the system density, or by varying the $s$-wave 
scattering length, e.g. by means of Feshbach resonances \cite{Inouye1998}: ultimately it is the product $g_{_{\rm 2D}} n_0$ which controls the effective soliton width. For numerical convenience, when probing different parameter regimes, we choose to fix $n_0$ and vary $g_{_{\rm 2D}}$ by increasing or decreasing the value of $a_s$ by up to three times its background value (see subsequent Fig.~3).

\item The maximum amplitude, $V_0$, of the density perturbation, which parametrizes the overall depth of the imprinted density notch, scaled to the gas chemical potential $\mu$.

\item The effective width of the density perturbation which (for a given $V_0$) is parametrised by $\sigma$.
\end{itemize}

Our subsequent generation and dynamical stability analysis is thus primarily based on the chosen control parameters $(V_0/\mu)$ fixing the depth of the emerging solitary wave excitations, and $(l_{\rm r}/\xi)$ setting the effective dimensionality of the system.

\section{Dynamics at $T=0$ \label{zero_temperature}}

In order to characterise the role of the relevant control parameters, and thus identify optimum regimes for solitonic generation in the idealised mean-field regime, we restrict our initial analysis to the (two-dimensional) Gross-Pitaevskii equation:
\begin{equation}
i\hbar\frac{\partial \psi({\bf r},t)}{\partial t}= 
\bigg(-\frac{\hbar^2}{2m}\nabla^2_{x,y}+ 
V({\bf r}) + g_{_{\rm 2D}}|\psi|^2-\mu \bigg)\psi({\bf r},t)
\quad\quad
\label{gpe}
\end{equation}

The idealised proof-of-principle generation of quasi-stable solitary waves is best demonstrated through a series of density and phase snapshots following the removal of the perturbing potential. Characteristic images are shown in Fig.\,\ref{initial_fig}c. More specifically, this figure shows (in situ) condensate density 
(top), renormalized (`carpet') density (middle) and phase (bottom) of the system 
at times  $t=36,43,72, 119,750\,{\rm{ms}}$ (left to right) after initiating the (ramped) perturbation, where $36\,{\rm{ms}}$ corresponds to the time that the perturbation is switched off, and $119$ ms and $750$ms the times after one and thirteen collisions.
The renormalized `carpet' plots are obtained in the usual way, by subtracting from the perturbed instantaneous density the
static density profile prior to the addition of the perturbation.
Throughout this work, density is given in units of $l_{0}^{-2}$, where $l_0=10\,{\rm{\mu m}}$
 is our reference `length unit'.

Figure\,\ref{initial_fig}c reveals\footnote{See also movie (2d.solitonic.long.evolution) in supplementary material.}
the emergence of counter-propagating sound waves moving rapidly away from the region of the density perturbation, 
followed by two slower counter-propagating structures of reduced density, which additionally feature a pronounced
phase slip across the density minima (bottom images).
Such generated structures propagate in opposite directions within the ring, 
collide with each other at the far end of the ring and emerge largely unaffected after the collisions,
as shown in the two rightmost frames of Fig.\,\ref{initial_fig}c. 
More specifically, for the case considered here (with $V_0=2\mu$),
the generated structures travel with a `mean' velocity $v \approx 0.5 c$ 
(based on the time taken to cross half the ring, i.e. $t\simeq66\,{\rm ms}$), 
where $c$ is the sound velocity in the medium, calculated at the peak density (i.e. at $r=r_0$). 
We will thus infer that such structures are solitary waves.
Further evidence for this is provided by their azimuthal 1D density cuts (subsequent Fig.\,\ref{azimuthal1.3}) revealing excellent agreement with the anticipated analytical 1D soliton solutions, for a soliton propagating at the same speed, with this observation broadly extendable also to the $T>0$ case (Fig.\,\ref{fig_1dsolitonall}).

The profiles shown here are relatively `clean', due to the gradual ramping of the perturbing potential, in stark contrast to equal duration potentials which are abruptly switched on and off, an example of which is shown in Appendix A.

The intensity of the laser is an important control parameter (for a given ramping on/off sequence), as it characterizes the maximum depth that the emerging nonlinear excitations can acquire. We anticipate requiring an intensity $V_0 \gtrsim \mu$, 
although we note that much higher values would imply that the two BECs become effectively disjoint in the region of perturbation (an effect identified in Ref.\,\cite{Burger2002} for elongated quasi-1D BECs). 
In our present work, we span intensity perturbations ranging from $V_0=0.5\,\mu$ to $V_0=10\,\mu$ within our ring trap geometry. 
As the symmetry of the Gross-Pitaevskii equation implies that
the two emerging structures are mirror 
images of each other, we focus here on one of the emerging structures, arbitrarily chosen here as the one propagating clockwise. 

\begin{figure}[!tbp]
  \centering
  \subfloat[]{\includegraphics[scale=0.3]{Gallucci_Fig2b.eps}\label{fig_velo_vs_intens}}
  \hfill
  \subfloat[]{\includegraphics[scale=0.4]{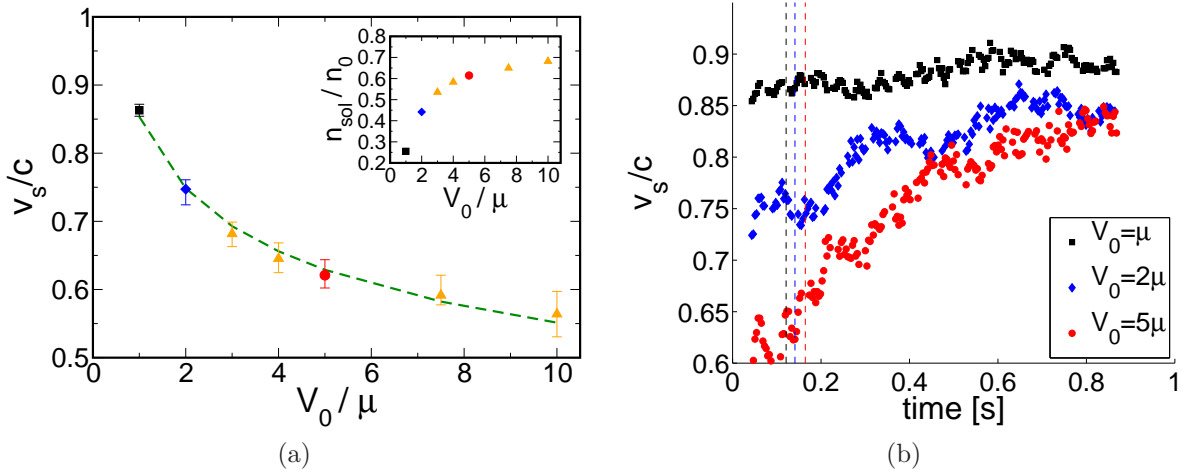}\label{fig_velocityratio}}
  \caption{(a) (Main Plot) Dependence of the (clockwise) soliton velocity, $v_s$,  scaled to the sound velocity, $c$ on (a) maximum amplitude $V_0$ of linearly-ramped perturbing potential (scaled to $\mu$) measured at the time when the solitary wave first reaches the top of the ring.
(Inset) Corresponding plot in terms of the soliton depth, $n_{\rm sol}$, scaled to the maximum unperturbed density $n_{\rm 0}$, obtained using the standard homogeneous relation 
$v_s/c=\sqrt{1-n_{\rm sol}/n_{\rm 0}}$ valid for purely 1D solitons \cite{kivshar1998}.
(b) Dependence of $v_s/c$ on time for different maximum amplitudes of the  
perturbing potentials
  $V_0/\mu=1\,({\rm black\,squares}),2\,({\rm blue\,diamonds}),5\,({\rm red\,circles})$.  All velocity ratios given here are based on the ratio of the instantaneous value of the soliton depth at $r_0$ (the radial distance specifying the location of the trap minimum) to the (peak) unperturbed density at that point.
The identifiable oscillations are likely due to the fact that the minimum of the solitary wave structure is not always located at $r_0$ [see also Fig.\,\ref{2dsoliton}]. We have also verified that 
the approximate determination of the soliton velocity based on measuring its motion around the ring yields similar results.
}
\end{figure}

As anticipated, higher values of $V_0/\mu$ lead to deeper (slower) solitary wave generation. To characterise this, Fig.\,\ref{fig_velo_vs_intens} shows the dependence of the effective azimuthal propagation speed (scaled to the local speed of sound measured at $r_0$) on the maximum amplitude of the perturbing potential.
As different structures travel at different speeds, and in order to avoid dependence on any initial excitations or related transient features (e.g. sound waves), we
have chosen to characterise the propagation speed at the point when the emerging solitary waves reach the top of the ring, i.e. around x=0.
We find $v_s/c \sim (V_0 / \mu)^{-\alpha}$, with a numerically-extracted exponent $\alpha \approx 0.18$. Using the standard expression for pure one-dimensional solitons in homogeneous settings \cite{kevrekidis_frantzeskakis_book_08}, we can re-write this formula in terms of the depth, $n_{\rm sol}$, of the soliton from the peak of the unperturbed density, as $n_{\rm sol} / n_0 \sim 1 - (V_0 / \mu)^{-2 \alpha}$, whose dependence is shown in the inset to Fig.\,\ref{fig_velo_vs_intens}.

Our numerical analysis indicates that our excitation scheme leads to the initial generation of highly excited nonlinear structures, which gradually evolve towards more robust structures which we shall henceforth refer to as `solitary waves'; nonetheless, such quasi-stable structures still feature some intrinsic dynamics.
Thus, our subsequent analysis is further
complicated by the fact that the emerging structures only approximately maintain their shape in time, in contrast to the case of a typical purely 1D soliton. Over longer timescales following the initial generation,
we find that the curvature and closed geometry of the ring trap, which imply that the solitary waves which feature their own internal dynamics are continuously accelerated in their circular motion towards/against each other, actually leads to their gradual decay. Such decay manifests itself in the usual form of  `anti-damping' \cite{Busch2000b}, i.e. growth of oscillation amplitude due to energy loss.
Although this decay rate is relatively slow, it does imply that the apparent depth (or equivalently speed) of the solitary waves decreases (increases) with time in a (semi)-monotonic way.
Figure\,\ref{fig_velocityratio} reveals some oscillations which could be attributed to a combination of the previously mentioned internal dynamics of the solitary waves, and their interactions with the propagating sound; the latter is somewhat reminiscent of (regular) oscillations induced by soliton-sound interactions in harmonically-trapped quasi-one-dimensional BECs \cite{Parker2003,Parker2010}. 
However, as evident from Fig.\,\ref{fig_velocityratio}, the decay of such structures in time is 
slow enough to allow for multiple collisions between the counter-propagating nonlinear structures, whose shape and speed appear to be only mildly perturbed by the collisions.
The time of revolution for each pair of counter-propagating solitary waves (i.e.\ each value of $V_0/\mu$) is shown by vertical dotted lines in Fig.\,\ref{fig_velocityratio}, with this figure spanning 5--7 revolutions.
This suggests that the observed anti-damping may not be a direct consequence of the (head-on) collisions, but rather a complicated effect due to a combination of the internally excited state of the solitary wave and the related azimuthal-radial mode coupling, in conjunction with the accelerated circular motion through the ring and the interaction with the propagating background sound excitations. 

To characterise the extent of such anti-damping numerically, we note that, after the emerging nonlinear excitations have completed  $\sim 7$ revolutions for the case $V_0=2\mu$ (intermediate blue diamonds in Fig.\,\ref{fig_velocityratio}), 
their velocity has increased by a mere $\sim 15\%$ compared to the initial value.
 This can be taken as concrete evidence supporting our interpretation of such structures as (slowly-decaying) solitary waves.

Having identified the potential for quasi-stable solitary-wave-like propagation in the ring, and the slow geometry- and interaction-induced underlying dissipation, there are two further main goals that we address in this work, namely the emergence of a regime where the solitary waves are reasonably stable and can be classified as ``solitonic'', and the role of thermal fluctuations.

Throughout this work,
 the quasi-2D nature of the system (fixed by $\mu<\hbar \omega_\perp$, or equivalently $l_\perp < \xi$) implies that transverse excitations outside the $(x,y)$ plane associated with 3D dynamical instabilities are suppressed. However, dynamical instabilities can also emerge in this two-dimensional geometry,
depending on the ratio of the effective ring width $l_{\rm r}$ to the healing length $\xi$, whose effect is discussed next (with the idealised limit of $l_\perp < l_{\rm r} < \xi$ corresponding to an effective 1D regime of practically stable solitonic propagation over experimental timescales).

\subsection{Solitonic Behaviour and Dynamical Instabilities}

\begin{figure}[!tbp]
\vspace{-2cm}
\centering
\includegraphics[scale=0.45]{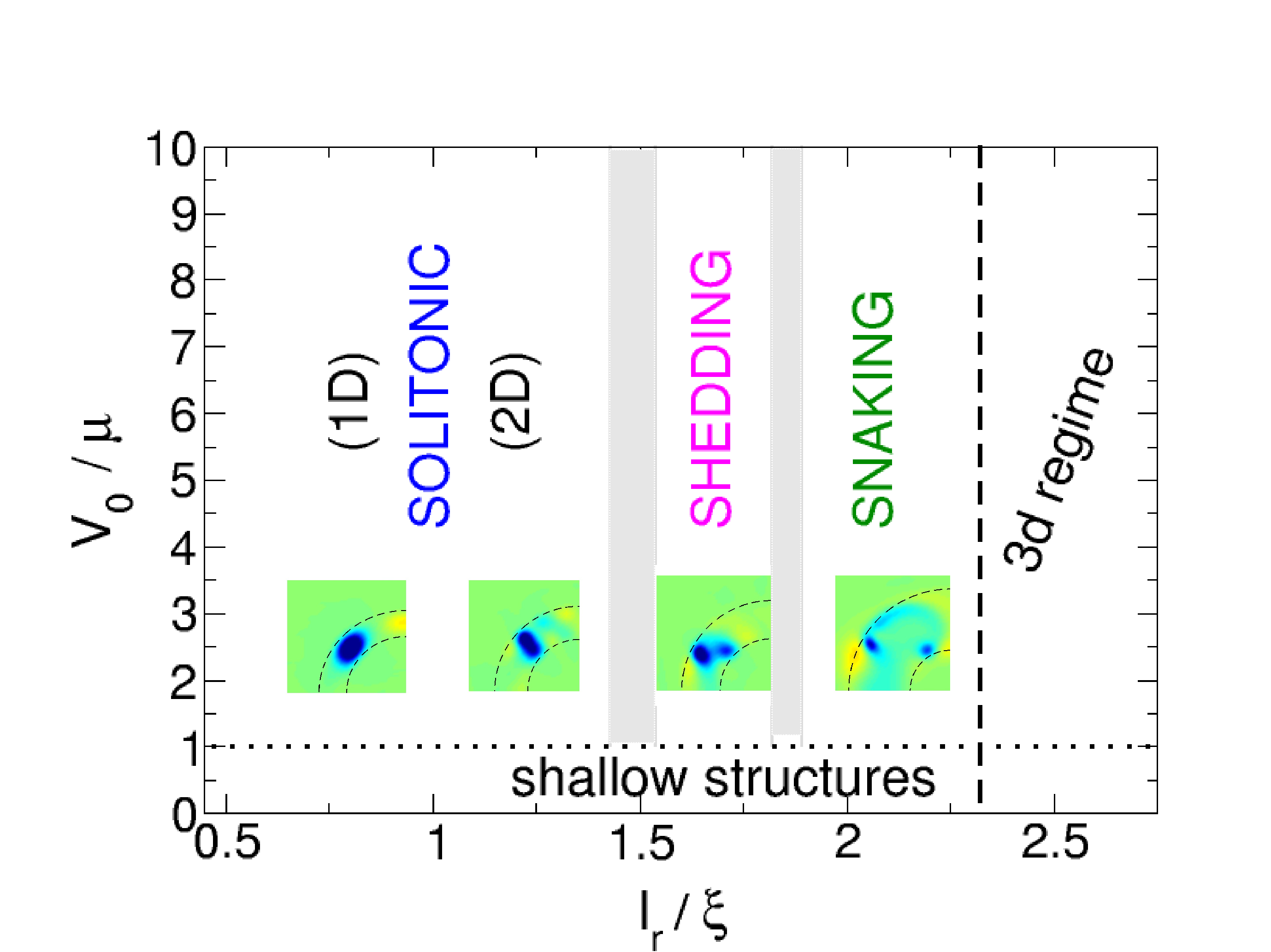}
  \caption{Phase diagram for the generation and stability of solitary waves, identifying 3 distinct regimes as a function of the dimensionless parameter 
  $l_{\rm r}/\xi$ for a broad range of values of 
 $V_0/\mu$ probed, separated by grey crossover regions.
The solitonic (stable) regime is separated from the snaking (unstable) 
  regime by an intermediate region (shedding) where the structures are highly excited and emit at least one pronounced density depression in their attempt to eventually maintain, after internal re-arrangement, some solitonic features.
 The characteristic behaviour defining each regime is displayed in the 2D carpet snapshots reported in each of those cases (for the clockwise propagating wave): 
  the dashed semi circles in each of these images indicate where the density drops to $10\%$ 
  of the peak equilibrium value. 
Crossover to the 3D regime (vertical dashed line) occurs roughly at $l_{\rm r}\approx2.3\xi$ (corresponding $l_\perp \approx \xi$).
The solitonic regimes features an internal `subdivision' around $l_{\rm r}=\xi$, with the 1D regime exhibiting perfectly symmetric solitons also in the radial direction [see subsequent Fig.\,\ref{2dsoliton}] and enhanced stability, facilitated by the suppression of radial excitations.
The horizontal dotted line at $V_0=\mu$ indicates the regime below which only rather shallow structures appear following the density perturbation, in the sense that the depth of the soliton (measured from the top) does not exceed $15\%$ of the peak unperturbed density, 
hence the soliton may not be pronounced enough to observe experimentally.
  }\label{phase_diagram}
\end{figure}

Figure\,\ref{initial_fig}c clearly demonstrates that 
quasi-stable solitary wave propagation is possible around the ring; however it is important
to further characterise such ensuing dynamics and identify regimes of rapid dynamical instabilities even in the 2D regime. By studying the dependence of the motion of the emerging solitary wave pairs on the ratio $(l_{\rm r}/\xi)$, we can identify 3 reasonably distinct dynamical regimes over the broad range $0 < V_0 / \mu < 10$ of density perturbation amplitudes probed, with the corresponding `phase diagram' shown in
Fig.\,\ref{phase_diagram}.
Specifically, there are two very distinct regimes, respectively associated with (quasi-stable) `solitonic' propagation and dynamically unstable `snaking' behaviour, gradually mediated by an intermediate regime that we have termed `shedding', 
due to the pronounced density emission from the stretched solitary wave. The insets 
to Fig.\,\ref{phase_diagram} illustrate characteristic snapshots identifying the key features of each of those regimes.

Note that propagating structures resembling dark solitons have also been observed (but not analysed in detail) in parallel recent 
work \cite{Wang2015_arxiv}, based on a similar experimental setup and perturbation scheme. In particular, numerical simulations 
reported in Fig.~8(left) of that paper show solitonic structures which appear to split into two,
suggesting this could be somewhat analogous to the behaviour observed by us in the `shedding' regime below.

In preparing the phase diagram (Fig.\,\ref{phase_diagram}), we have chosen to probe the distinct regimes by varying the value of ($l_{\rm r}/\xi$) for fixed $l_{\rm r}$.
We choose to control the size of the healing length, $\xi$, by changing the value of the scattering length from its background value, $a_s$.
In so doing, we have decided to work with a {\em constant peak density} of $2500\,l_0^{-2}=25$ atoms$/{\mu m}^2$, which facilitates an easier comparison of the different emerging density profiles, rather than fixing the total atom number; in turn, this implies also adjusting the value of the chemical potential $\mu$. More specifically, for the probed regime $0.8<l_{\rm r}/\xi<2.3$, the scattering length spans the range
$\approx [0.4 a_s \, , \, 3 a_s]$ (i.e.\ $1.1\,{\rm nm}<a_s<8.3\,{\rm nm}$ for $^{23}$Na used here), 
while the chemical potential still satisfies the 2D criterion through the condition $0.25<\mu/\hbar\omega_\perp<1$ (with the number of atoms lying in the range $\approx[12000 , 28000]$ respectively). 
We have also verified that changing $\omega_\perp$ (instead of $a_s$) in the range $\approx [0.2 \omega_\perp \, , \, 9.2 \omega_\perp]$, while still keeping the peak density $n_{_{\rm 2D}}(r_0)\propto (\xi^2 a_s \sqrt{\omega_\perp})^{-1}$ fixed to the same constant value, yields the same physics. 

\begin{figure}[!tbp]
  \centering
  \hspace{1cm}
  \subfloat{\includegraphics[scale=0.51]{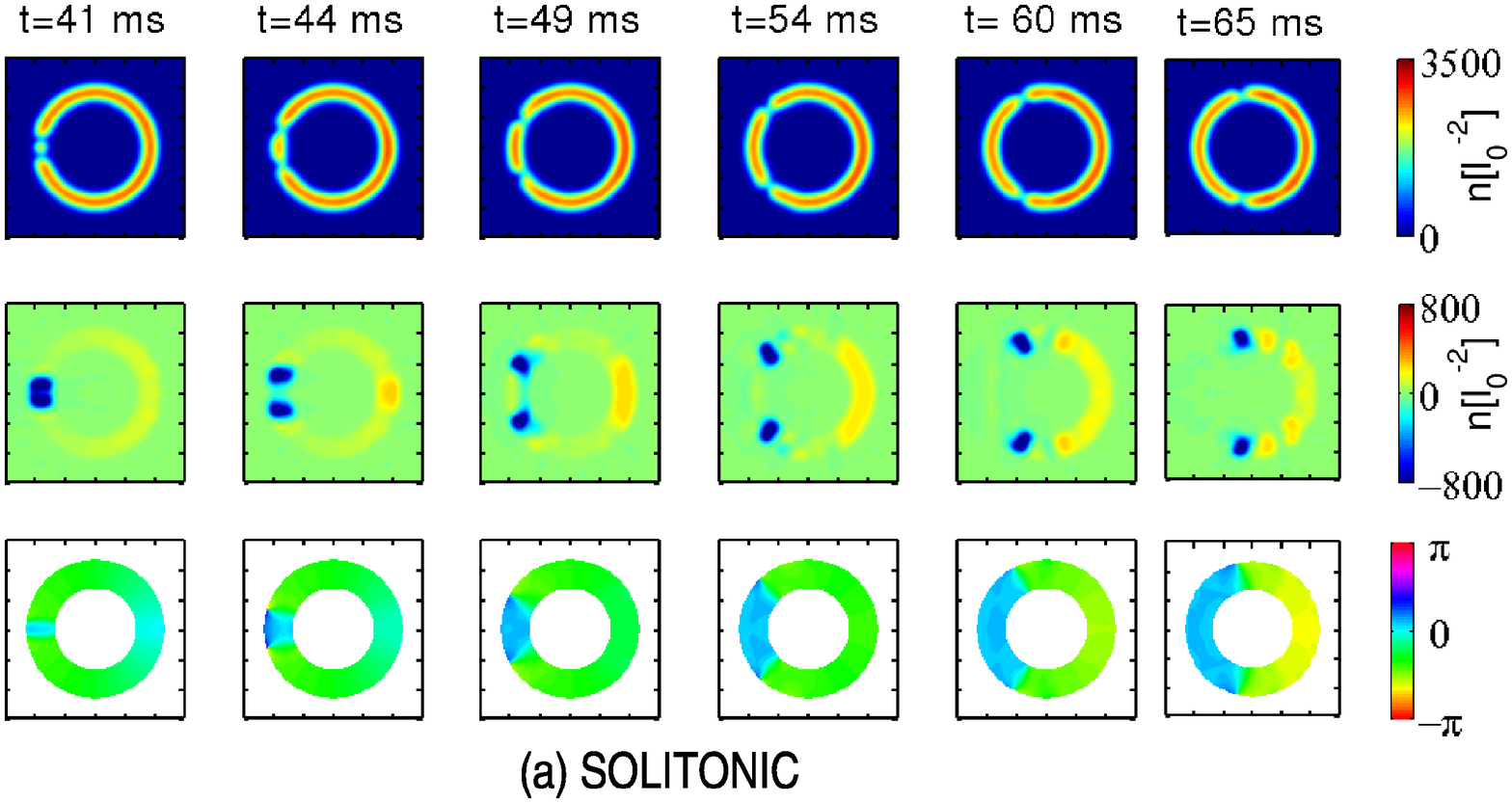}\label{solitonic}}\\
  \hspace{1cm}
  \subfloat{\includegraphics[scale=0.51]{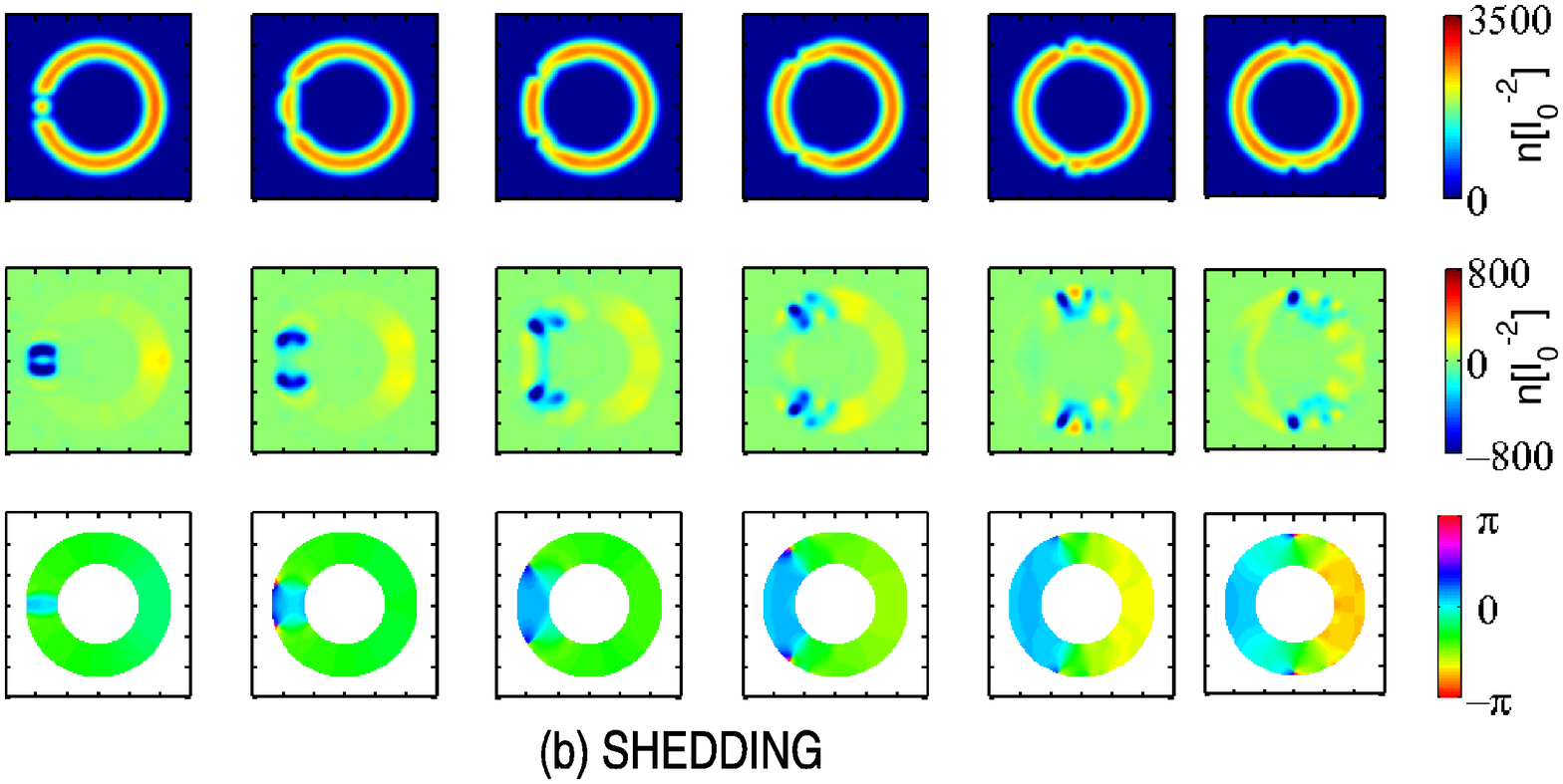}\label{shedding}}\\
  \subfloat{\includegraphics[scale=0.51]{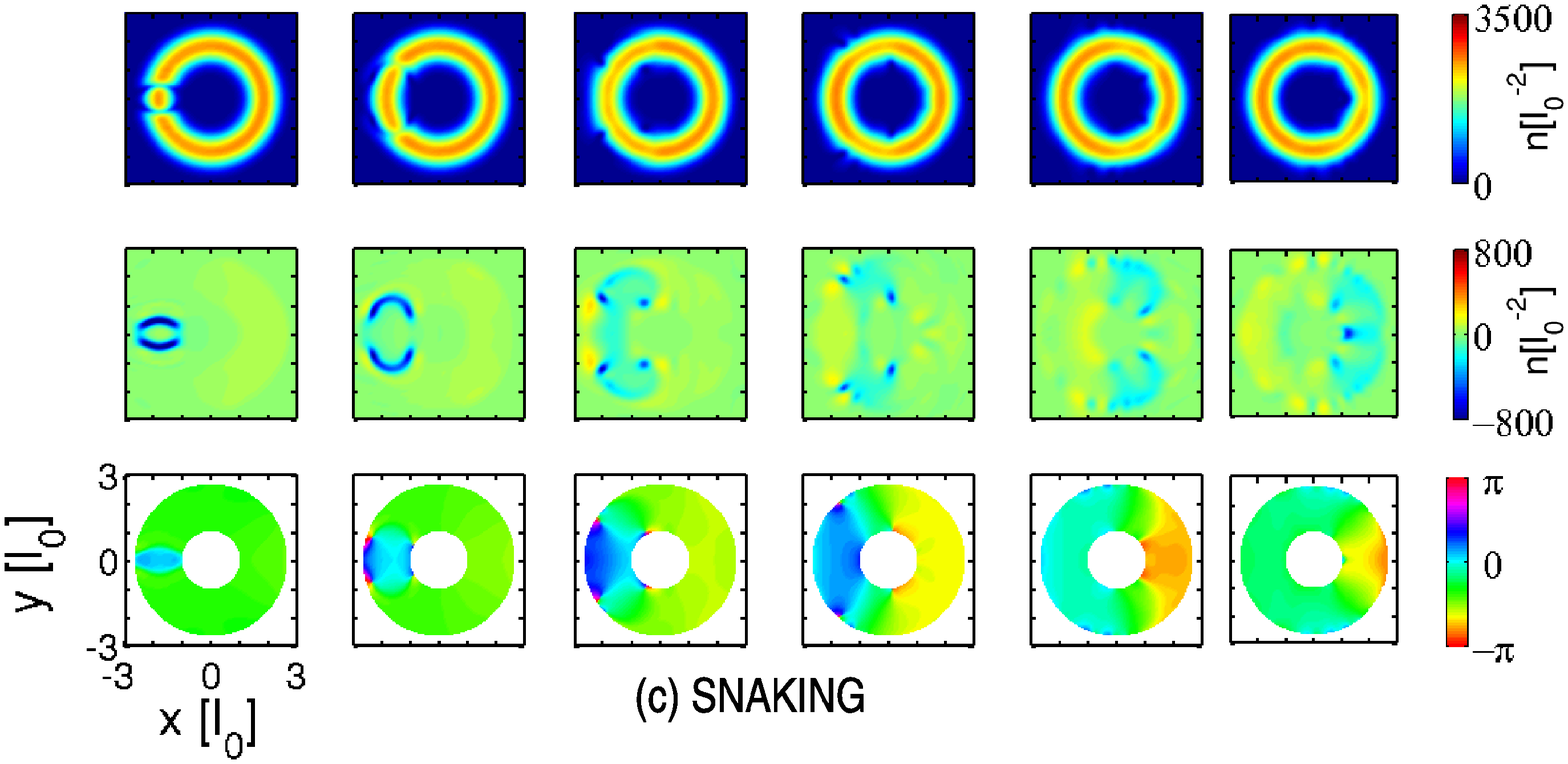}\label{snaking}}
  \caption{Comparison between density, carpet and phase profiles at times 
  $t=41,44,49,54,60,65\,{\rm{ms}}$ (left to right) for the (2D) `solitonic' (a), `shedding' (b) and 
  `snaking' (c) regimes, where $l_{\rm r}/\xi=1.3,1.5,2.2$ respectively. 
Phase plots include, as before, a phase mask for densities lower than $10\%$ 
  of the peak equilibrium density.}
\end{figure}

The distinct regimes identified, separated by a crossover rather than a sharp boundary, are discussed further below, with reference to 
Fig.\,4 showing detailed snapshots of the evolution of set times for all cases\footnote{See supplementary material for movies in corresponding regimes (1d.solitonic, 2d.solitonic, shedding, snaking).}.\\

\noindent {\bf `Solitonic' regime} {\em (Fig.\,\ref{solitonic})}: this regime, an example of which was discussed in Fig.\,\ref{initial_fig}c, is characterised by the propagation of quasi-stable solitary waves which extend practically over the whole width of the ring, and are thus classified here as `solitonic'.
The example demonstrated in Fig.~2 revealed a gradual decay of the emerging solitons, through a slow decrease in their depth, with their overall structure remaining largely unaffected. Such structures do however still exhibit some internal dynamics, mainly associated with  coupling between radial and azimuthal degrees of freedom with increasing values of $l_{\rm r}/\xi$.

One can also internally sub-divide this regime, based on the value of $(l_{\rm r}/\xi)$, since for values $l_{\rm r} < \xi$ one arrives at an effectively 1D geometry satisfying $l_\perp<l_{\rm r}<\xi$.

\begin{figure}[!tbp]
  \centering
  \subfloat[]{\includegraphics[scale=0.4]{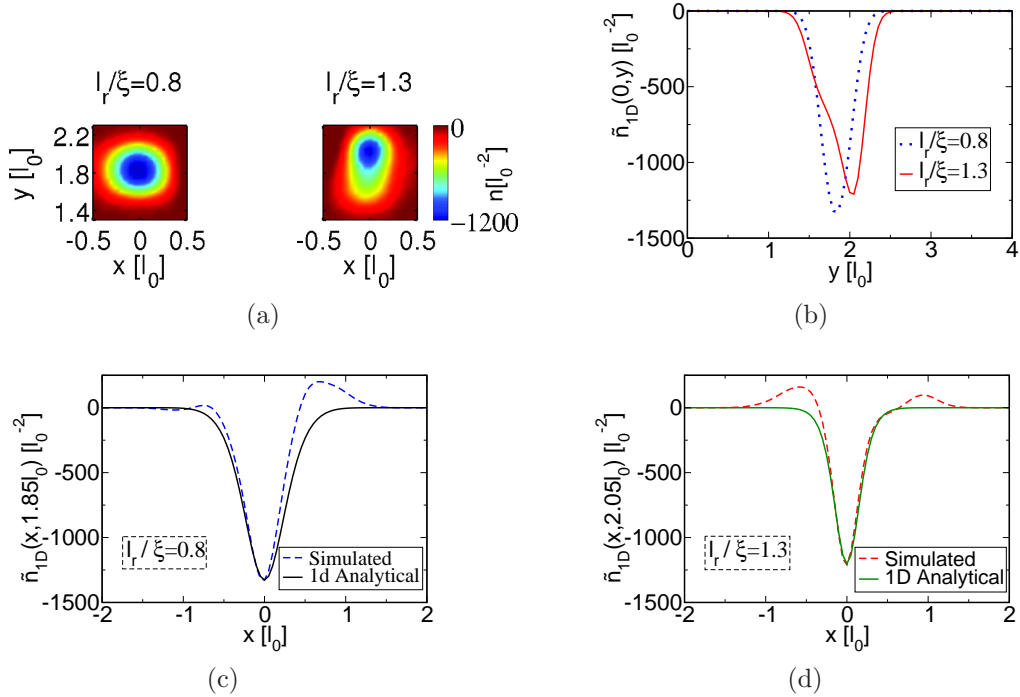}\label{2dsoliton}}
  \hspace{1cm}
  \subfloat[]{\includegraphics[scale=0.2]{Gallucci_Fig5b.eps}\label{radial}}\\
  \subfloat[]{\includegraphics[scale=0.2]{Gallucci_Fig5c.eps}\label{azimuthal0.8}}
   \hspace{2cm}
  \subfloat[]{\includegraphics[scale=0.2]{Gallucci_Fig5d.eps}\label{azimuthal1.3}}
  \caption{(a) 2D carpet plot showing a zoomed in image of the clockwise solitary wave
  when it first passes through the top of the ring (i.e. at $x=0$)
  for the cases $l_{\rm r}/\xi=0.8$ (left) and $1.3$ (right).
(b)--(d) Corresponding one-dimensional radial and azimuthal density slices $\tilde{n}_{\rm 1D}$. Specifically:
(b) Comparison of 1D radial profiles (at $x=0$) revealing that for higher values  of $l_{\rm r}/\xi$,
  the density minimum, $n_{\rm min}$, does not occur at the point where the trap reaches its minimum, but slightly
  shifted towards the outer edge of the ring.
(c)--(d) Comparison  between simulated
  1D azimuthal carpet profiles (dashed lines) and corresponding 1D solitonic analytical solution (solid line) for the two cases, with
density cuts taken at $y_0=1.85\,l_0$ and $2.05\,l_0$,
  for $l_{\rm r}/\xi=0.8$ and $l_{\rm r}/\xi=1.3$ respectively.
Analytical soliton solutions are constructed from their measured speed through the relation
$v_s/c=\sqrt{(n_{\rm min}(0,y_0)/n_{\rm 0}(0,y_{0}))}$, where $n_{\rm min}$ is the soliton depth and $n_{\rm 0}$ the unperturbed equilibrium density, with the healing length calculated at the peak unperturbed density.
}
\end{figure}

A comparison of density profiles between the case with $l_{\rm r}/\xi=0.8$ and our reference case of Fig.\,\ref{initial_fig}c ($l_{\rm r}/\xi=1.3$) is shown in Fig.\,\ref{2dsoliton}.
Figure\,\ref{2dsoliton} shows 2D carpet plots of the solitonic structures when they are at the top of the ring, also plotting corresponding one-dimensional density cuts in the 
 radial (Fig.\,\ref{radial}) and azimuthal (Fig.\,\ref{azimuthal0.8}--d) directions.
We find that, although for $l_{\rm r}<\xi$ the solitonic structures are symmetric along the radial direction about the middle of the ring, increasing the ratio $l_{\rm r}/\xi$, makes the structures less symmetric 
around the minimum of the ring trap (Fig.\,\ref{2dsoliton}--b), with 
the location of the density minimum shifted towards the outer edge of the trap; 
more specifically, by comparing the radial profiles (Fig.\,\ref{radial}), 
while for $l_{\rm r}/\xi=0.8$ the density minimum lies exactly at the point 
where the trap reaches its minimum $(x=0,y=1.85\,l_0)$, in the case of $l_{\rm r}/\xi=1.3$  
the minimum occurs at $(x=0,y=2.05\,l_0)$ instead, which is presumably related to the excited solitonic dynamics seen in its subsequent evolution.
 
The 1D azimuthal profiles for each case are also shown in Fig.\,\ref{azimuthal0.8}--d. Comparing these to the anticipated analytical 1D soliton profile \cite{Morgan1997,Reinhardt1997,Busch2000b} for the same speed, we find excellent agreement, thus fully supporting our claim that such structures can be termed `solitonic'.
The restriction of radial excitations significantly enhances the solitonic stability, as further illustrated in Appendix B.
\noindent {\bf `Shedding' regime} {\em (Fig.\,\ref{shedding})}: this is an intermediate regime in which the emerging nonlinear structures display pronounced internal dynamics at early times. 
A defining characteristic in this regime is that, following the removal of the perturbing potential, the initial azimuthal stretching of the
propagating density depressions is balanced by a pronounced density re-arrangement, which results in the gradual separation of a significant density wave from the main depression, with the emitted density wave eventually dispersing:
in some cases (smaller values of $l_{\rm r}/\xi$), the remaining structure partially `recovers' towards a more shallow `solitonic' profile spanning 
a significant fraction of the width of
the entire ring radially, which is however less stable than those in the identified `solitonic' regime; 
in other cases (higher values of $l_{\rm r}/\xi$) the width of the remaining solitary-wave excitation remains clearly less than the width of the ring. In both cases, such structures continue moving azimuthally (even if they only span a fraction of the radial ring width), and such `solitary waves' appear to still survive multiple collisions.
The emitted density waves can also be thought of as secondary shallower solitary waves, and their respective initial depths (and thus survival lifetimes) are increased with increasing $(V_0/\mu)$, with the crossover region satisfactorily accounting for such behaviour.
As the values of $l_{\rm r}/\xi$ increase beyond a certain thereshold region, the emerging structure stretches so much that it actually bends and breaks due to the background density gradient, in a manner reminiscent of the snaking instability observed e.g. in optical media \cite{Mamaev1996} or elongated BECs \cite{Anderson2001}.\\

\noindent {\bf `Snaking' regime} {\em (Fig.\,\ref{snaking})}: the nonlinear structures emerging after the removal of the perturbation 
deform so substantially along the ring, 
becoming dynamically unstable, as the azimuthal width of stretched density depressions greatly exceeds the healing length $\xi$, implying that solitary wave solutions can no longer be the lowest energy states of the system.
Each of the two counter-propagating nonlinear structures then breaks into two 2D vortices 
(representing the planar mapping of 3D vortex rings), located near the inner and outer edges of the ring trap. Such dynamics is highly reminiscent of the observed `snaking instability' in which 3D dark solitary waves decay into vortex rings \cite{brand2002,kevrekidis_frantzeskakis_book_08,Proukakis2004.job,Shomroni2009}.

Having investigated in reasonable detail the role of the various `geometrical' control parameters for the optimal generation of solitonic structures in ring-trap BECs, we now briefly address the important role of temperature and fluctuations on the form and lifetime of the emerging solitonic structures.


\section{Dynamics at \, $T>0$ \label{finite_temperature}}

Temperature can be introduced into the Gross-Pitaevskii model in two closely related ways, by the controlled addition of fluctuations into the numerical simulations \cite{Proukakis_ICPress,Proukakis2008,blakie_bradley_08,brewczyk_borowski_04}. In the simplest approach (Sec.\,\ref{sec.sgpe_gpe}), we start with an appropriately thermalised initial state, at some temperature $T$, described by a fluctuating classical field which is then propagated  by the usual Gross-Pitaevskii equation [Eq.\,(\ref{gpe})].
This approach is typically referred to as the `classical field' method \cite{kagan_svistunov_92a,Davis2001,Berloff2002,brewczyk_gajda_07,blakie_bradley_08} 
(being closely related to the finite temperature truncated Wigner \cite{Steel1998,Sinatra2001,proukakis_schmiedmayer_06,Cockburn2011b}), and relies on the ergodicity of the Gross-Pitaevskii equation.
Such a model has been used to study, among other phenomena, spontaneous soliton generation \cite{witkowska_deuar_11} and dark soliton stability \cite{Martin2010,martin_ruostekoski_10a}.

A more complete treatment of fluctuations requires both time-dependent stochastic (noise) fields and a dissipation term (with the two related through a fluctuation-dissipation relation \cite{Stoof1999}). In this case, both fluctuations and dissipation arise from the coupling of the stochastic classical field, representing the low-lying, highly-occupied `classical' modes of the system up to a cutoff to higher-lying (thermal) modes. The addition of the dissipation implies that the system relaxes (with a rate dictated by $\gamma$) to the equilibrium set by the heat bath parameters (temperature $T$ and chemical potential $\mu$).

Both approaches can atually appear as different limits of the stochastic Gross-Pitaevskii equation (SGPE) \cite{Stoof1999,Stoof2001,Duine2001}, which in our current 2D setting takes the form
\cite{Cockburn2012}:
\begin{equation}
i\hbar\frac{\partial \phi({\bf r},t)}{\partial t}= 
\big[1-i\gamma\big] \bigg(-\frac{\hbar^2}{2m}\nabla^2_{x,y}+ 
V({\bf r}) + g_{_{\rm 2D}}|\phi|^2-\mu \bigg)\phi({\bf r},t)+ \eta({\bf r},t),
\quad\quad
\label{eq_sgpe}
\end{equation} 
where $\phi({\bf r},t)$ now represents the multi-mode stochastic `classical' field cumulatively describing the low-lying 
modes of the Bose gas (see also the closely-related Stochastic Projected Gross-Pitaevskii Equation \cite{Gardiner2003,blakie_bradley_08}).
This should be directly contrasted to the usual Gross-Pitaevskii equation [Eq.\,(\ref{gpe})], where $\psi({\bf r},t)$ denotes simply the {\em condensate} wavefunction.
In Eq.\,(\ref{eq_sgpe}), (thermal) fluctuations are mimicked by the presence of the noise term $\eta({\bf r},t)$ which has Gaussian 
correlations of the form $\langle\eta^*({\bf r},t)\eta({\bf r'},t')\rangle=
2 \hbar \gamma k_{\rm B} T \delta({\bf r}-{\bf r'}) \delta(t-t')$, where $\gamma$ parametrises the strength of the noise and damping.

Although such an equation should be solved numerous times with different stochastic fields, with the results appropriately averaged, one can actually attribute an indirect physical interpretation to each numerical trajectory, as representing a plausible experimental run.
For a discussion of the usefulness of single stochastic trajectory analysis and how to extract meaningful averaged parameters from this, see e.g.\ our earlier work on stochastic dark soliton dynamics in harmonic traps \cite{Cockburn2010}.

In order to investigate the effect of temperature on the soliton dynamics, we focus on our (largely sound free, 2D) reference case of Fig.\,\ref{initial_fig}c, for which relatively deep solitons were clearly visible in the $T=0$ limit.
We now use those two different limits of the SGPE to discuss the ensuing soliton motion at finite temperatures, through indicative single-trajectory results.

%


\subsection{`Classical Field' Method \label{sec.sgpe_gpe}}

\begin{figure}[!tbp]
  \centering
  \subfloat[$t=0\,{\rm ms}$]{\includegraphics[scale=0.49]{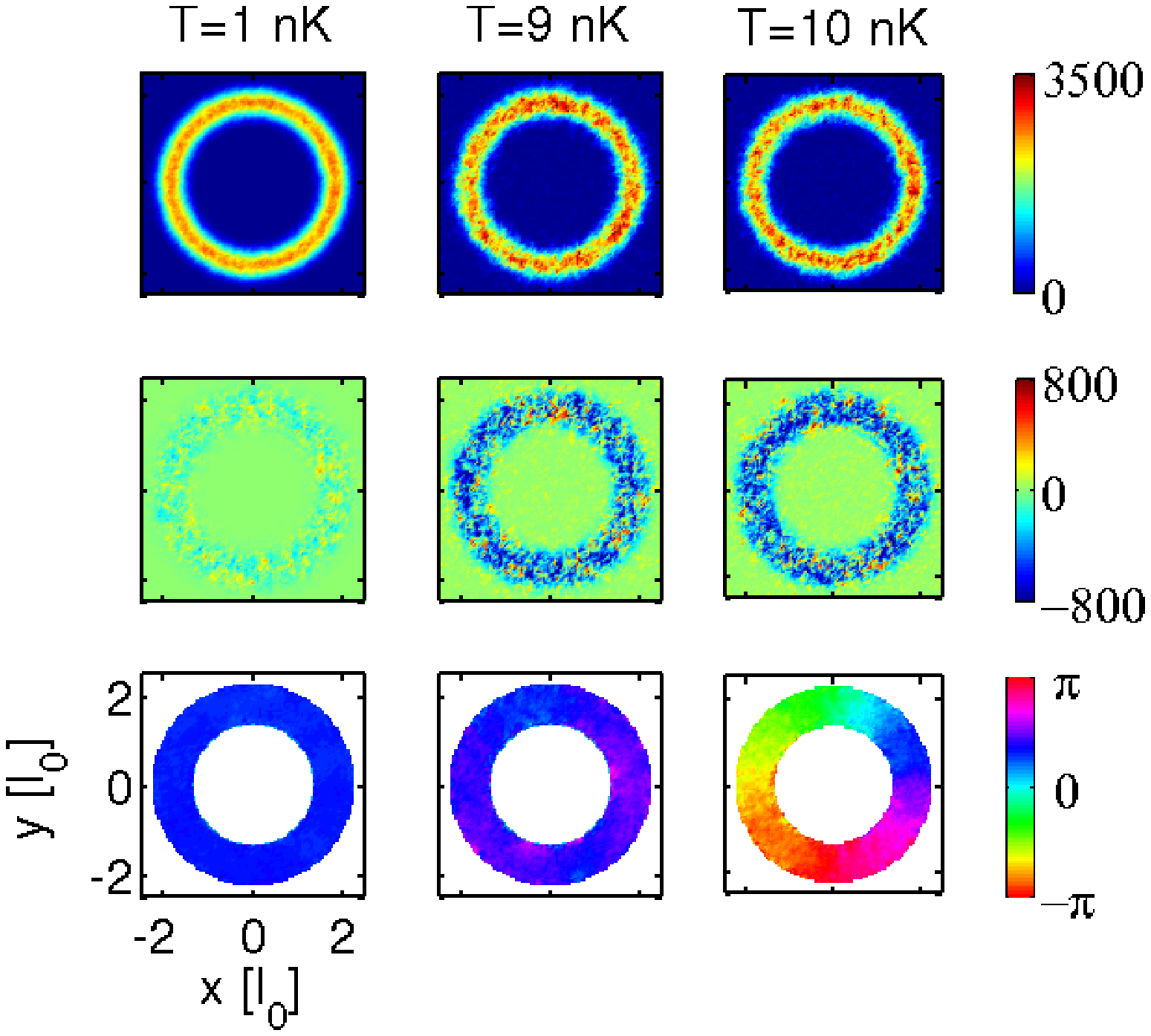}\label{fig_sgpe1}}
  \hfill
  \subfloat[$t=238\,{\rm ms}$]{\includegraphics[scale=0.49]{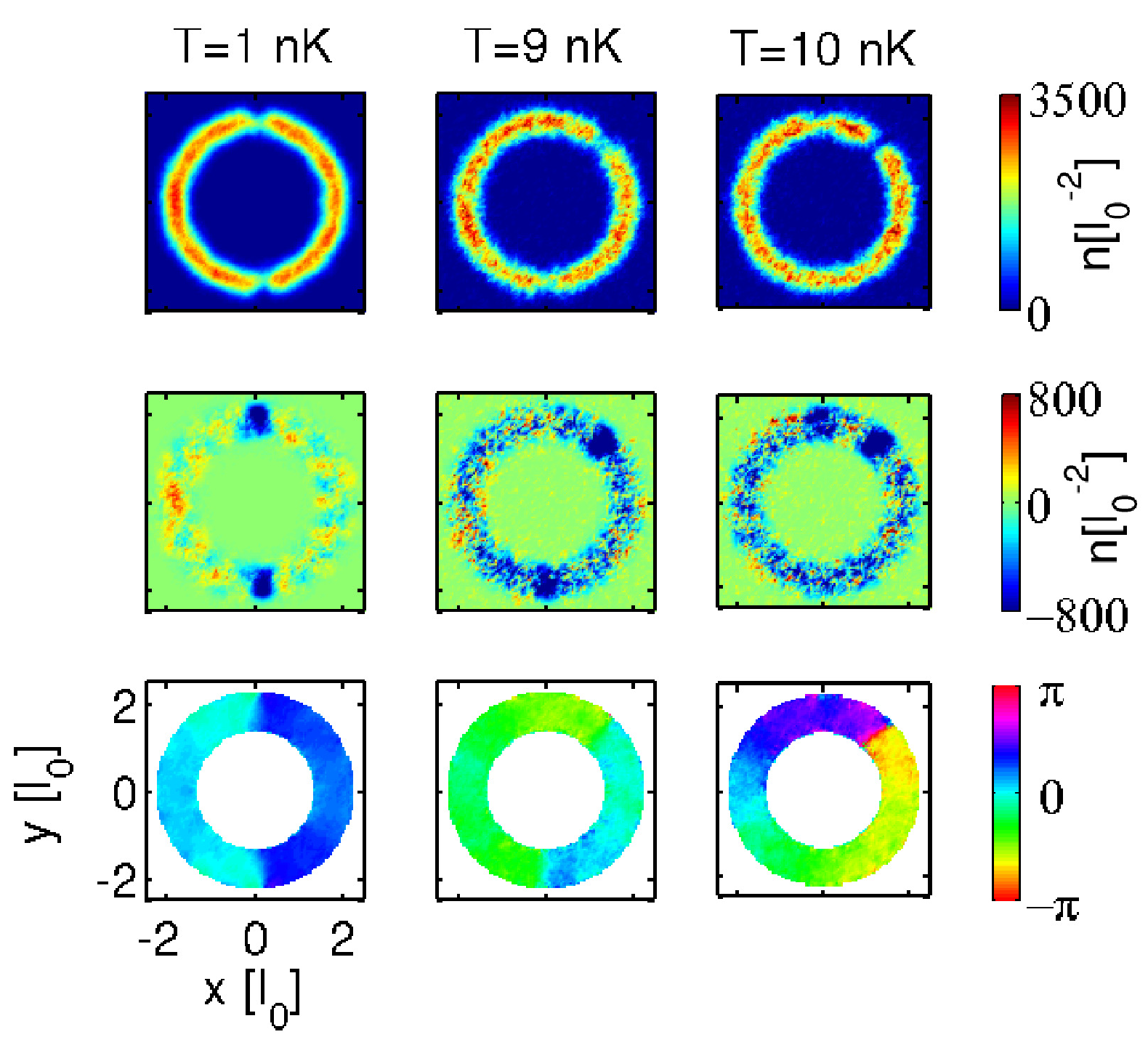}\label{fig_sgpe2}}
  \caption{(a) Initial fluctuating equilibrium states at temperatures 
  $T=1, 9, 10\,{\rm nK}$ %
  [respective atom numbers $N=15910, 18776, 18902$]. 
(b)   Post-perturbation evolution at time $t=238\,{\rm ms}$
 when the solitons have each undergone just over one and a half revolution of the ring, having thus already interacted 
  three times (except in the $T=10$\,nK case, where the two solitons have only interacted twice). 
Note that these plots show the entire classical field density, $|\phi|^2$, rather than the condensate density, $|\psi|^2$, of the ordinary Gross-Pitaevskii equation shown until now. One could in principle perform further analysis to extract the corresponding density images for the (quasi-)condensate, which would look smoother; however, the location and nature of the solitons in the condensate would closely mimic the effects seen in the classical field plots, adding no further insight into the soliton stability. Moreover, the displayed noisy profiles are closer in nature to what would be observed in an experiment.
}
\end{figure}

In this section we generate the `initial' state, i.e.
state prior to adding the density perturbation,
as an appropriate thermal noisy equilibrium state, via {\em dynamical equilibration} of the Stochastic Gross-Pitaevskii Equation [Eq.\,(\ref{eq_sgpe})].
After equilibration, we switch off both dynamical noise and dissipation, which amounts to propagating our noisy thermalised initial state 
via the {\it ordinary} Gross-Pitaevskii equation.
Representative images of equilibrium density profiles in the presence of fluctuations are shown\footnote{See also movie in accompanying material [2d.solitonic.T9nK].} in Fig.\,\ref{fig_sgpe1} for different temperatures of the unperturbed thermal state. As before, we show
densities (top), renormalized densities or `carpet' plots (middle) and phase (bottom) plots for $T=1, 9, 10\,{\rm nK}$ 
(left to right).
The carpet plots (middle) are generated by subtracting 
from the single stochastic run perturbed density at a given time the 
corresponding $T=0$ (mean field) unperturbed equilibrium result.
As expected, the fluctuations in the background density increase with increasing temperature. 

A few comments are in order here:

(i) The stochastic numerical evolution leads to the generation of a different 
random phase in each numerical simulation, such that the underlying phase differs from run to run and temperature to temperature. 
To facilitate a more direct comparison of the soliton dynamics between the different 
temperature cases, we therefore numerically eliminate the {\em initial} random 
phase difference (i.e. the phase difference of the equilibrium configuration prior 
to turning on the perturbation) among the cases $T=1$ and $9\,{\rm nK}$ shown here.  

(ii) The $T= 10$ nK case we have chosen to show here is slightly different, as it contains a persistent current (here with a winding number 1) at our $t=0$ time labelled as `equilibrium', which is simply a reflection that the system has not yet actually fully equilibrated. 
This persistent current has appeared here spontaneously during our equilibration process (and will eventually decay after a sufficiently prolonged evolution).
The reason for this appearance can be traced back to our method of generating the initial state, which is actually based on dynamical equilibration following an instantaneous numerical quench, a process which is known to support such spontaneous defect formation, in accordance with the Kibble-Zurek mechanism \cite{Das2012}. We have also checked that over numerous simulations we get a distribution of both flow-free solutions and persistent currents with positive and negative winding numbers, including also higher winding numbers, in qualitative agreement with experiments \cite{Corman2014}. Persistent currents can also be generated in lower temperature cases, so our choice of displaying the 10 nK case here with a persistent current is because it yields a clean persistent current over strong background density fluctuations, thus providing clear evidence of a strikingly different motion around the ring, combining both persistent currents and density fluctuations.
Note that 10 nK is also the highest (optimal) temperature we can realistically probe in our setup, to avoid atoms populating transversally excited modes, which are not accounted for in our purely 2D scheme. 
While instructive to show how the presence of the persistent current affects the 
generation/propagation of the solitary waves, and although we could ensure that 
the perturbing potential is added after the persistent current decays \cite{Rooney2013}, we have 
chosen not to investigate this case further, since experiments aiming 
to generate solitons would also choose initial conditions without 
an intrinsic flow pattern.

(iii) As our simulations have been done at constant chemical potential, 
the atom numbers increase slightly with increasing temperature (up to 20\%), 
but we do not expect this to have a significant effect on our presented analysis
(other than, e.g., in making the speed of sound slightly temperature-dependent.)

The evolution of density and phase on these noisy initial 
states after the addition and removal of the perturbing potential
is shown in Fig.\,\ref{fig_sgpe2}. 
In all cases, analogously to the $T=0$ case, we can detect two emerging structures which tend to 
propagate in opposite directions ($T=1, 9 \,{\rm nK}$) and appear to remain 
largely unchanged through their collisions. 
The profiles shown here are taken after the solitons have already undergone 
one and a half revolution (time $t=238$\,ms), such that they
have met each other and interacted three times (at $x \approx 1.85 l_0$, $x 
\approx -1.85 l_0$ and again at $x \approx 1.85 l_0$).

We have performed a study based on numerous individual classical field simulations, based on completely random initial conditions (generated through SGPE equilibration), and observe a range of features commented upon below:

As the temperature (represented by thermally-induced background density inhomogeneity) 
increases, the motion of the two counter-propagating solitons reveals small 
differences (although the mean propagation speed remains {\em approximately} constant). 
We can attribute this to a combination of two effects 
(largely guided here by our earlier work on dark solitons \cite{Cockburn2010}): 
on the one hand, the presence of random fluctuations in the initial state, implies 
that the emerging dark solitons are not identical; moreover, even though the average 
noise amplitude at each temperature is fixed, at any time each soliton is nonetheless 
propagating through a different random noisy background configuration, which 
introduces small random `kicks' to the soliton motion around the ring. As a result, 
the two solitons do not collide exactly at $y=0$, and at any given time their 
respective positions are not exactly mirror-images of each other [see e.g. $T=9$\,nK 
case in Fig.\,\ref{fig_sgpe2}].
The fact that the generated structures appear in the same location after three 
consecutive interactions (and having done more than one full revolution in the ring) 
strongly suggests that the solitonic nature of such structures persists even in the presence of initial fluctuations.

Interestingly, we do not find a systematic net effect of temperature, i.e. the average position of the solitonic waves after few revolutions  (averaged over $\sim 10$ stochastic runs), is only mildly perturbed from the corresponding $T=0$ case without displaying a clear dependence on temperature.
This is a feature that we have also observed in our previous work on dark soliton dynamics in purely one-dimensional geometries
\cite{Cockburn2010,Cockburn_PhD}. This appears to be in partial disagreement to the findings of Refs.\,\cite{Martin2010,martin_ruostekoski_10a}, where it has been argued that dark solitons propagating on an initially fluctuating background exhibit some decay\footnote{In Refs.\,\cite{Martin2010,martin_ruostekoski_10a} the Truncated Wigner approximaton is used for the quasi-condensate description; quantum and thermal fluctuations are retained in this approach, and the quasi-condensate is obtained by using an extension of the Bogoliubov theory to treat low-dimensional Bose gases \cite{Mora2003}.
In our model the equilibrium solution is determined self-consistently via the SGPE, and contains information about both density and phase fluctuations; the phase-coherent condensate, or suppressed density-fluctuations quasi-condensate could then be extracted \textit{a posteriori} from the SGPE classical field, and its density is expected to qualitatively resemble the plotted classical field density, but with the fluctuations largely suppressed.}, 
whereas our work provides evidence of temperature-dependent modulations, but no net decay.
 While our analysis does not give (or intend to give) a conclusive answer to this issue, it does suggest that if classical field simulations correctly predict the soliton dynamics, then multiple soliton collisions should routinely arise in carefully engineered experiments.

The presence of the persistent current in the third ($t=0$) and sixth (post-density-engineering) subplots of Fig.\,6  imparts an additional 
flow velocity to the two solitons, thereby significantly speeding the motion of 
the co-flowing soliton, while simultaneously decelerating the soliton travelling 
against the flow. As a result, in the presence of a persistent current, the two 
solitons exhibit a net relative speed between them, and the motion in this case 
deviates significantly from that when no persistent current is present, 
where the mean soliton's  
$x$-coordinates were found to be approximately equal.

\begin{figure}[!tbp]
  \centering
  \subfloat[]{\includegraphics[scale=0.45]{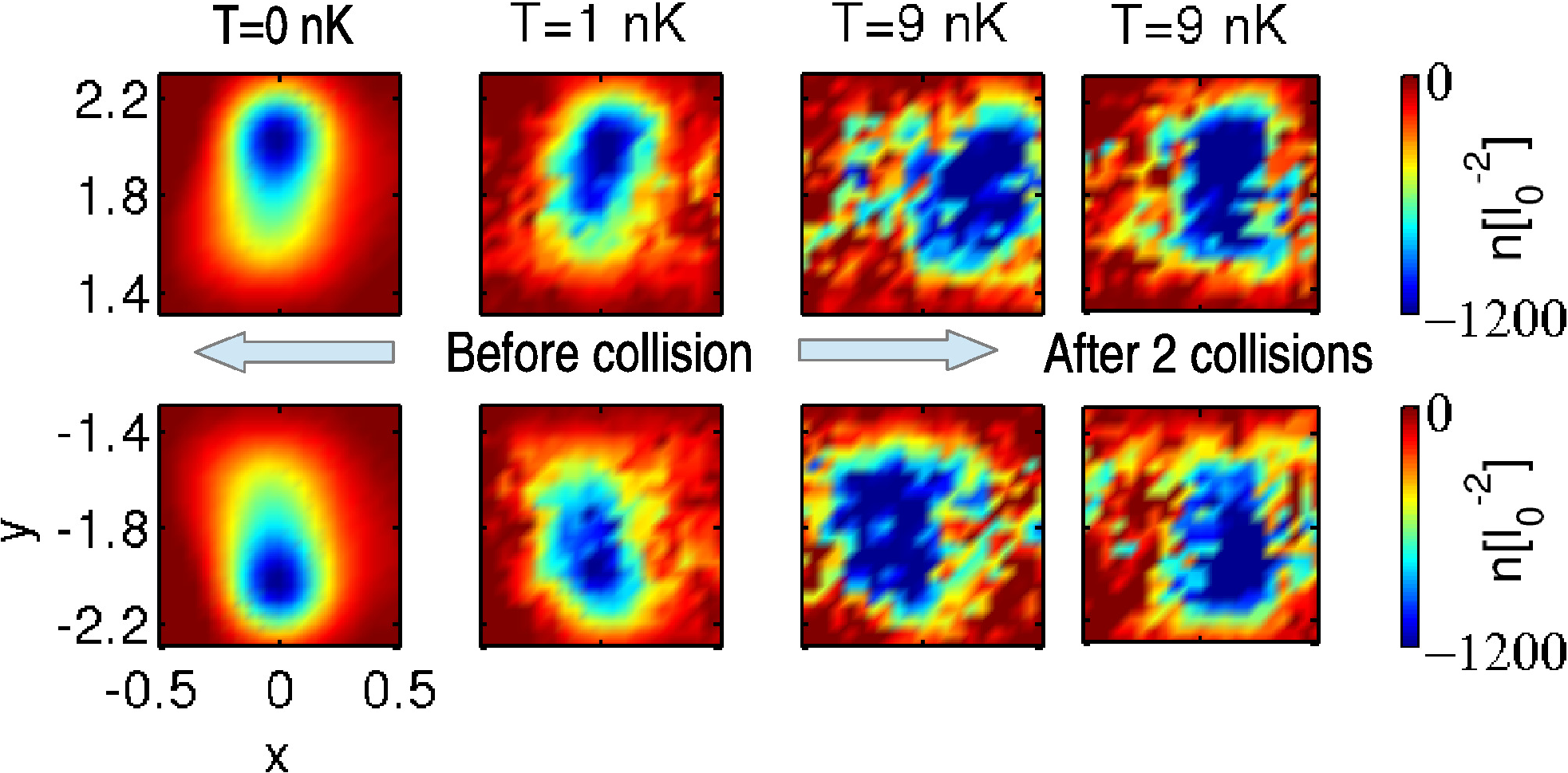}\label{fig_2dsolitonall}}
  \hfill
  \subfloat[]{\includegraphics[scale=0.22]{Gallucci_Fig7b.eps}\label{fig_1dsolitonall}}
  \caption{ (a) (Left 3 Columns) Temperature dependence of clockwise (top) and counterclockwise (bottom) solitary waves at $T=0,1,9\,{\rm nK}$ (from left to right) at the top of the ring (prior to their first interaction); (Right Column) Corresponding solitary wave images at T=9\,{\rm nK} after the waves have interacted twice. (b) Comparison between 1D azimuthal renormalized density profiles for the clockwise solitary waves of Fig.\,\ref{fig_2dsolitonall} (top) [at $y=2\,l_0$] before the first interaction.
Note that the resolution of these images is sub-$\mu$m, implying that an experimental study would actually reveal smoother profiles, in agreement with our earlier experimental-theoretical comparisons \cite{Cockburn2011a,Cockburn2012}.
} 
\end{figure}

The role of fluctuations on the soliton density is illustrated in Fig.\,\ref{fig_2dsolitonall} which compares its form in the absence ($T=0$) or presence ($T=1$, 9 nK) of background fluctuations in the initial state. Although the actual position of the soliton in an individual numerical run jitters about the mean equilibrium position, and its profile becomes less well defined due to the underlying fluctuations, the fluctuations themselves do not appear to critically affect the underlying solitonic shape even after a few collisions [see rightmost image in Fig.\,\ref{fig_2dsolitonall}].
To verify the solitonic nature of the emerging structures prior to any collisions, Fig.\,\ref{fig_1dsolitonall} plots their azimuthal one-dimensional density cuts slightly after their generation (when located at the top of the ring) for finite temperatures, contrasting them to the pure $T=0$ case. This figure clearly shows that although the fluctuations noticeably modify the density profiles, the underlying solitonic nature reflected by the central width of the density depression set by the healing length $\xi$ remains clearly visible.

Based on all above findings, we would thus argue that the solitonic nature appears to persist both in the initial and dynamical regimes, when modelling the non-equilbrium soliton dynamics on top of a fluctuating {\em initial} state.

\subsection{Full stochastic evolution \label{sec.sgpeonly}}

To further improve on our earlier $T>0$ predictions, we now consider the dynamics resulting from density engineering in the context of the dynamical SGPE, in which the classical modes of the system described by $\phi({\bf r},t)$ exhibit full dynamical coupling to the high-lying modes of the system, i.e. maintaining here both {\em dynamical noise} $\eta({\bf r},t)$ and dissipation $\gamma$.

\begin{figure}[!]
\centering
\includegraphics[scale=0.6]{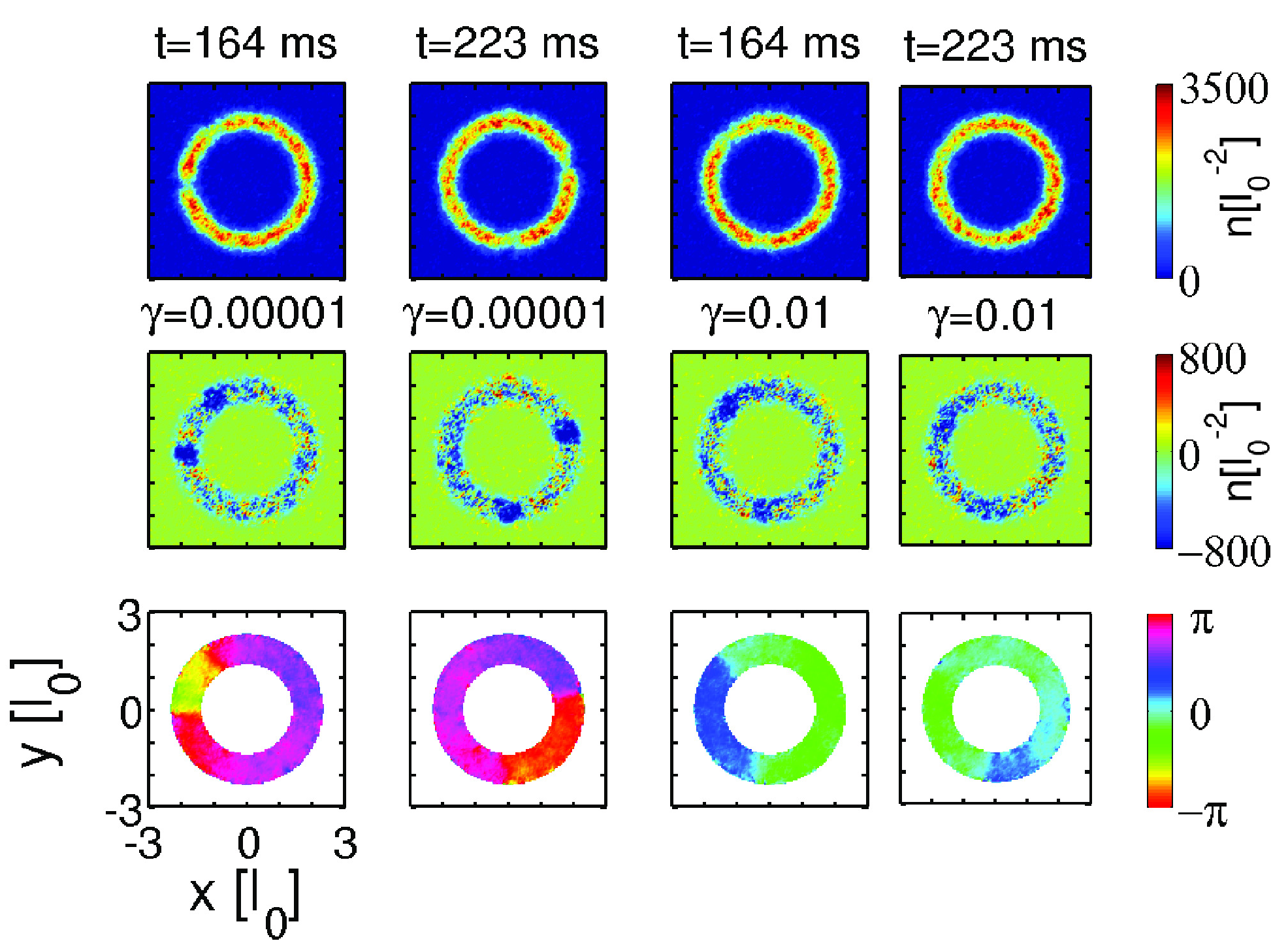}
  \caption{Comparison between density (top), carpet (middle) and phase (bottom) profiles at T=9\,{\rm nK} for $\gamma=10^{-5}$ (two left columns) and $\gamma=10^{-2}$ (two right columns) at time $t=164,223\,{\rm ms}$, corresponding roughly to one and two revolutions completed respectively.}\label{fig_fullsgpe}.
\end{figure}

The presence of $\gamma$ ensures the system eventually relaxes to an equilibrium profile dictated by the heat bath $T$ and $\mu$, thus clearly leading to the gradual decay of any generated excitations, at a rate directly dependent on $\gamma$. Although $\gamma$ is often considered a `phenomenological parameter', an analytical prediction for its value does exist 
\cite{Penckwitt2002,duine_leurs_04,Cockburn_PhD,Proukakis2008,blakie_bradley_08}.
Importantly, recent work with the closely-related Stochastic Projected Gross-Pitaevskii Equation (SPGPE) 
demonstrated excellent agreement between theoretical predictions based on the theoretically-predicted $\gamma$ value and experimental findings, in the context of persistent current decay in a ring trap \cite{Rooney2013}, thus suggesting that such simple analytical estimates yield reasonably realistic values.

Using the predicted analytical expression \cite{Penckwitt2002,Rooney2013}, leads in our system to an estimated value of $\gamma \sim 10^{-5}$, which we use here simply as a guide. Figure\,8 (left two images) shows snapshots of the post-density-engineering evolution of the solitonic structures, revealing the persistence of clearly-identifiable solitonic structures even after 2 full revolutions (or 4 mutual collisions).
Given the somewhat crude estimated values for the decay parameter $\gamma$, the two rightmost plots of Fig.~8 show the corresponding case with a much larger (heuristically chosen) value of $\gamma\sim10^{-2}$. Even in this case, which features enhanced soliton decay, we still find evidence of the (attenuated) solitonic structures surviving after at least one full revolution around the ring ($t=164$ms, corresponding to two collisional events), as shown in the third set of plots in Fig.~8.

We thus conclude that although thermal excitations can significantly perturb the shape and reduce the lifetime of the solitonic excitations, their presence and collisions could be observable under realistic experimental conditions, provided the temperature is not too high. This is in qualitative agreement with previous discussions of soliton stability in elongated 3D harmonically-trapped BECs \cite{Jackson2007}.

%

\section{Conclusions}
 
We have investigated the conditions under which the addition of a carefully-engineered 
density perturbation to an atomic Bose-Einstein condensate contained within a ring trap can controllably generate pairs of counter-propagating solitonic excitations, demonstrating that such structures should
in fact survive (multiple) collisions and revolutions around the ring, even at finite temperatures.

Optimum experimentally-relevant conditions for their observation include tight transverse confinement along the direction orthogonal to the plane of the ring (denoted by the frequency $\omega_{\perp}$) and small atom number (or equivalently chemical potential $\mu$) such that the two-dimensional condition $\mu < \hbar \omega_\perp$ is satisfied, thus suppressing three-dimensional dynamical instabilities. Nonetheless, the azimuthal and radial degrees of freedom can still couple with each other, and a form of dynamical `snaking' instability was found to persist even in two-dimensional geometries ($l_\perp<\xi<l_{\rm r}$), unless the effective radial ring length $l_{\rm r}$ satisfied $l_{\rm r} \lesssim 1.5 \xi$, where $\xi$ denotes the healing length of the gas determining the soliton width.
Although experimentally challenging, a further reduction in the radial width of the ring trap, such that $l_{\rm r} < \xi$, would lead to an effectively one-dimensional geometry, significantly stabilizing the soliton against dynamical decay. An alternative, perhaps more easily accessible way to achieve the same 1D dimensional reduction, could be based on reducing the scattering length by means of a Feshbach resonance. In the particular realistic geometry discussed throughout this paper, a reduction in the scattering length of $^{23}$Na by a factor of 2.5 from its background value was sufficient to generate stable one-dimensional solitonic structures over the probed regime of numerous collisions.

To better distinguish the solitonic nature of the excitations over other (linear/background) excitations, we found it advantageous to use a density engineering protocol in which the intensity of the perturbing laser beam is gradually turned on over a period of few tens of ms, reaching a maximum intensity of few times the chemical potential.
To simpify the ensuing dynamics, and the observation of the propagating solitonic structures, it is advantageous to only generate a single counter-propagating soliton pair, which requires the waist of the laser beam to be narrow, broadly comparable to the healing length.

Looking at the role of thermal effects in the quasi-two-dimensional regime $k_{\rm B}T \lesssim \hbar \omega_\perp$, we performed an analysis based on two complementary models commonly used for non-equilibrium soliton dynamics (classical field simulations and stochastic Gross-Pitaevskii equation). Despite their somewhat distinct predictions, both models consistently indicated a high likelihood of observing solitonic generation, azimuthal propagation, and occurence of (possibly a few) solitonic collisions under realistic experimental conditions and temperatures.

We thus hope that our study will assist experimentalists in engineering quasi-stable solitonic propagation in closed ring-trap circuits, and that such nonlinear excitations could in the future prove useful for atomtronic applications.

\section{Acknowledgments}
We are grateful to Mark Edwards for originally suggesting this project to us, and to Bill Phillips for pointing us towards density engineering in the early stages of the development of this project.
Funding was provided by the UK EPSRC (Grants No. EP/I019413/1 and EP/K03250X/1), and we also acknowledge discussions with Andrew Baggaley, Carlo Barenghi, Simon Gardiner, Mariella Loffredo and Nick Parker.

Data supporting this publication is openly available under an 'Open Data Commons Open Database License'. Additional metadata are available at: 10.17634/122626-1. Please contact Newcastle Research Data Service at rdm@ncl.ac.uk for access instructions.

\appendix
\section*{Appendix A: Role of instantaneous and broad density perturbations}
\setcounter{section}{1}

In the main text we have argued that efficient sound-free generation of a single counter-propagating solitonic pair requires, in addition to the other carefully considered parameters, a gradual excitation scheme, and a narrow laser beam. To highlight the importance of those additional control parameters, here we give (for fixed other parameters) evidence of the post-perturbation dynamics when either of those criteria is not satisfied.

\begin{figure}[t]
  \centering
  \subfloat[]{\includegraphics[scale=0.397]{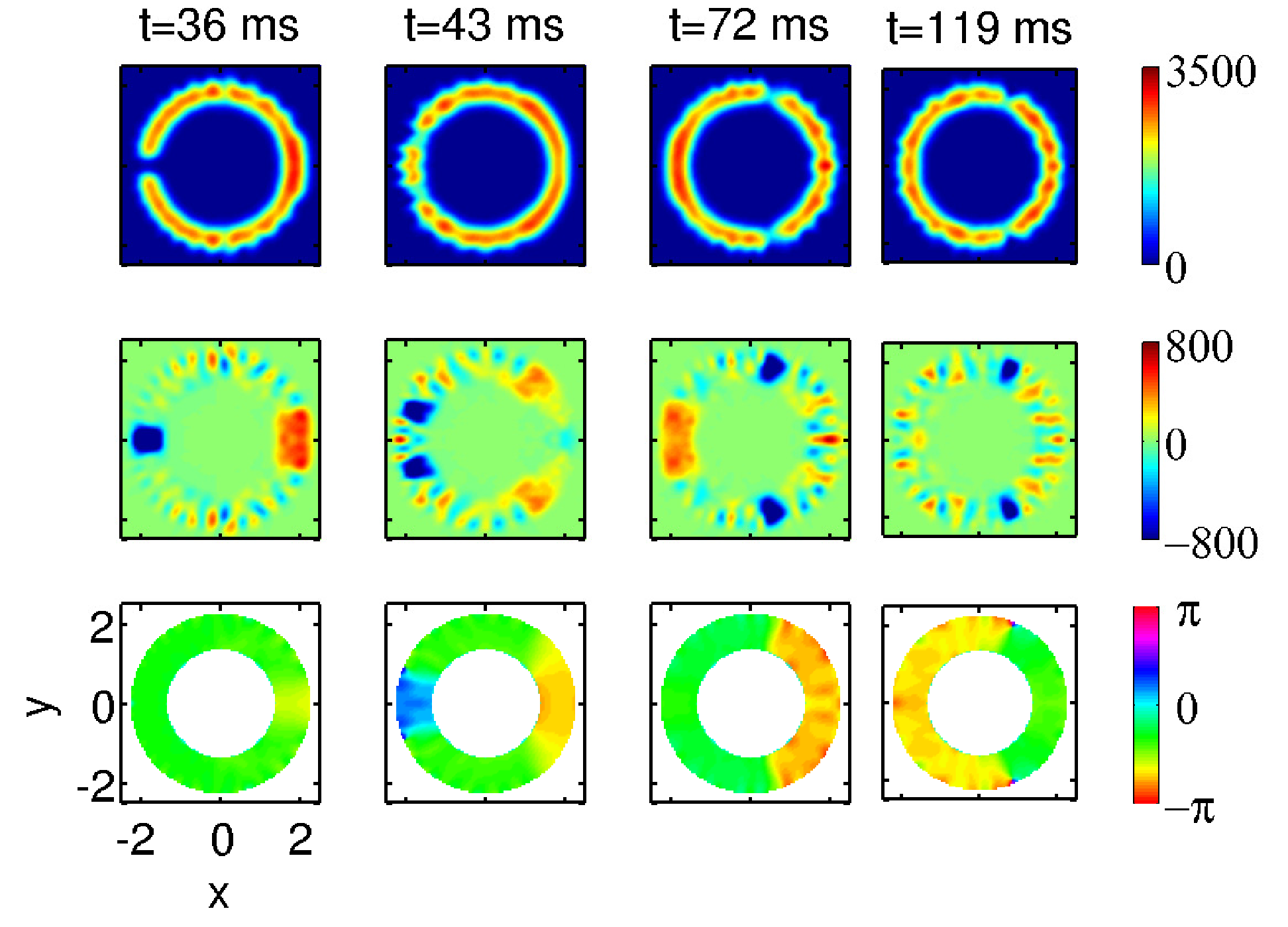}\label{fig_sudden}}
  \hfill
  \subfloat[]{\includegraphics[scale=0.397]{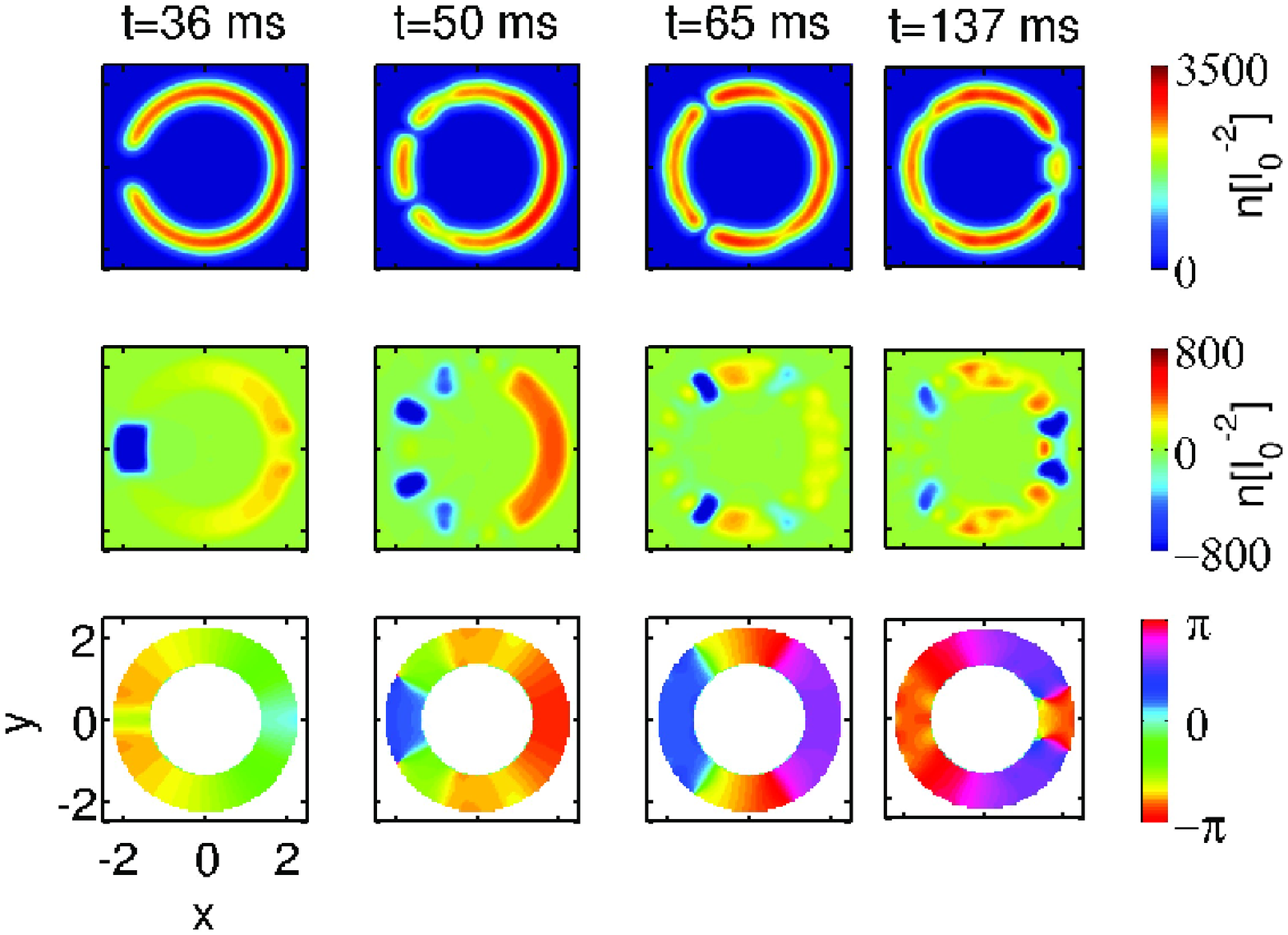}\label{fig_largesigma}}
  \caption{Evolution of emerging solitonic structures for different density engineering protocols: 
(a) As in Fig.\,\ref{initial_fig}c, but with the perturbation added suddenly 
(i.e. over a time equal to our time unit $\sim36\,{\rm{\mu s}}$),
shown here at the moment the perturbation is switched off ($t=36$ms) and subsequent times  $t=43,72, 119\,{\rm{ms}}$ (left to right) when the system evolves freely. 
(b) As in Fig.\,\ref{initial_fig}c (so with a ramped perturbation), but with $\sigma/\xi\approx 1.5$; these are shown here at slightly different times 
$t=36, 50, 65, 137\,{\rm{ms}}$ 
(left to right) in order to best reveal the two ensuing solitonic pair dynamics.}
\end{figure}

Figure\,\ref{fig_sudden} shows 
the situation analogous to our $T=0$ reference case, when the Gaussian perturbation is suddenly turned on (over a physical timescale $\sim 36 \mu$s corresponding to our time discretization unit), depicting again the ensuing density and phase dynamics at the same times as in Fig.~1c. More specifically, after being turned on, the perturbing potential is here kept at the constant maximum value of $V_0=2\mu$ for 36ms, before being again `instantaneously' removed.
A detailed comparison of Fig.~A1a and Fig.~1c reveals that
the turning on/off sequence therefore does not appear to significantly modify the details (depth/speed) of the emerging solitonic structure, but rather it controls the amount of emitted sound during the generation 
process, which in turn indirectly affects the 
long-term soliton evolution due to soliton-sound interactions.

Figure\,A1b shows the effect of increasing $\sigma/\xi$ to the value 1.5, which is here shown (for $V_0=2 \mu$) to lead to the 
eventual dynamical generation of more than one pair of counterpropagating dark 
solitons (of different depths).
We have checked that for values of $\sigma$ up to the value 
of $\xi$ (such that the Gaussian perturbation half-width at $1/e^2$ is $\sim 2 \xi$), 
a single pair of counterpropagating solitons is generated, 
placing an effective limit on experimental perturbing potentials able to generate 
only single (as opposed to multiple) dark soliton pairs.

\appendix
\section*{Appendix B: Solitonic Propagation in 1D versus 2D Regimes}
\setcounter{section}{2}

\begin{figure}[b]
\centering
\includegraphics[scale=0.5]{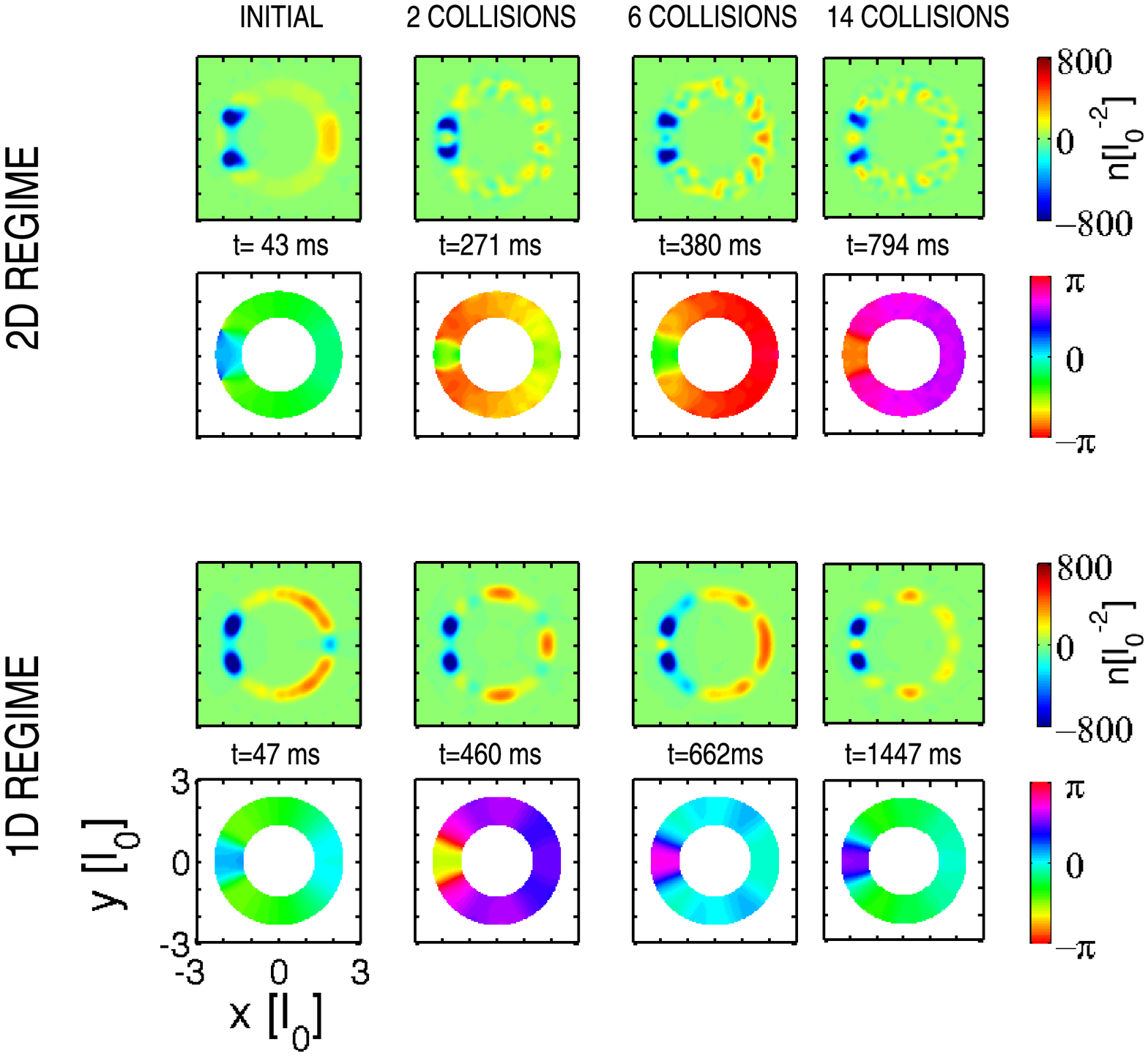}
  \caption{Typical carpet (top rows) and phase (bottom rows) snapshots depicting the initial generation (leftmost column) and subsequent propagation (after the number of indicated collisions) of the counter-propagating solitonic structures in the 1D ($l_\perp<l_{\rm r}<\xi$) and 2D ($l_\perp<\xi<l_{\rm r}$) solitonic regimes at the indicated times. As the generated solitons have different speeds, and the snapshots have been chosen to depict times when the solitons have returned to their initial position after a certain number of collisions (number of revolutions is half the number of collisions), the actual times of those snapshots do not coincide in the 1D and 2D cases. Parameters as in Fig.~1c, except in the 1D regime where $l_{\rm r} = 0.8 \xi$ (facilitated through the use of the modified $0.4a_s$ scattering length.)
  }\label{compare_lrxi_0.8_to_1.1}.
\end{figure}

For completeness, we present in Fig.~\ref{compare_lrxi_0.8_to_1.1} a comparison of typical snapshots of the long-term evolution in the solitonic regime between the 1D and 2D limits, as characterised by the parameter $l_{\rm r}/\xi$. Shown here are images shortly after the solitonic structures are first engineered (left columns), and then when the solitons have returned to the same position after undergoing two, six, and fourteen collisions (corresponding to one, three and seven revolutions around the ring respectively). This clearly demonstrates the robustness of the solitonic structures against collisions, while also showing the much more confined nature of the excitations in the 1D regime ($l_\perp<l_{\rm r}<\xi$), for which all radial excitations are completely suppressed. 
As a result, any solitonic excitations in 1D happen along the ring (structures occasionally appear more `oval-shaped' than circular), with our numerics indicating no noticeable change in the soliton speed/depth over the probed timescales (other than a small oscillation in their respective values). This is in contrast to the 2D regime (top images), where the solitons, although still reasonably robust to collisions, do exhibit changes in their profiles in time (exhibiting a coupling between azimuthal and radial degrees of freedom), and also gradually decay (albeit at a rather slow rate). While the 1D regime evidently provides optimal conditions for observing such an effect, our simulations indicate that the main effects should still be largely visible even when $l_{\rm r}$ slightly exceeds the healing length.
\section*{References}

\bibliography{Gallucci_et_al.bib}
\bibliographystyle{iopart-num.bst}

\end{document}